\documentclass[modern,trackchanges]{aastex631}

\usepackage[normalem]{ulem} 
\usepackage{rotating}
\usepackage{graphicx}

\def\beginrotate{\begin{Sbox}}
\def\endrotate{\end{Sbox}\rotatebox{-90}{\TheSbox~~}}

\usepackage{booktabs}
\newcolumntype{N}{>{\centering\arraybackslash}m{.5in}}
\newcolumntype{G}{>{\centering\arraybackslash}m{2in}}

\shorttitle{A Survey of CO, CO$_2$, and H$_2$O in Comets and Centaurs}
\shortauthors{Harrington Pinto et al.}
\graphicspath{{./}{figures/}}

\begin{document}

\title{A Survey of CO, CO$_2$, and H$_2$O in Comets and Centaurs}

\correspondingauthor{Olga Harrington Pinto} \email{oharrington@knights.ucf.edu}

\author[0000-0002-2014-8227]{Olga Harrington Pinto}
\affiliation{Department of Physics, University of Central Florida}

\author[0000-0003-4659-8653]{Maria Womack}
\affiliation{Department of Physics, University of Central Florida}
\affiliation{National Science Foundation}

\author{Yanga R. Fernandez}
\affiliation{Department of Physics, University of Central Florida}

\author{James Bauer}
\affiliation{University of Maryland}




\begin{abstract}

\label{sec:abstract}

CO and CO$_2$ are the two dominant carbon-bearing molecules in comae and have major roles in driving activity. Their relative abundances also provide strong observational constraints to models of solar system formation and evolution but have never been studied together in a large sample of comets. We carefully compiled and analyzed published measurements of simultaneous CO and CO$_2$ production rates for 25 comets. Approximately half of the comae have substantially more CO$_2$ than CO, about a third are CO-dominated and about a tenth produce a comparable amount of both. There may be a heliocentric dependence to this ratio with CO dominating comae beyond 3.5 au. Eight out of nine of the Jupiter Family Comets in our study produce more CO$_2$ than CO. The six dynamically new comets produce more CO$_2$ relative to CO than the eight Oort Cloud comets that have made multiple passes through the inner solar system. This may be explained by long-term cosmic ray processing of a comet nucleus's outer layers. We find (Q$_{CO}$/Q$_{H_2O}$)$_{median}$ = 3 $\pm$ 1\% and (Q$_{CO_2}$/Q$_{H_2O}$)$_{median}$ = 12 $\pm$ 2\%. The inorganic volatile carbon budget was estimated to be (Q$_{CO}$+Q$_{CO_2}$)/Q$_{H_2O}$ $\sim$ 18\% for most comets. Between 0.7 to 4.6 au, CO$_2$ outgassing appears to be more intimately tied to the water production in a way that the CO is not. The volatile carbon/oxygen ratio for 18 comets is C/O$_{median}$ $\sim$ 13\%, which is consistent with a comet formation environment that is well within the CO snow line.

\end{abstract}


\keywords{comets, solar system formation}


\section{Introduction} \label{sec:intro}


The physico-chemical conditions in the protosolar nebula varied with distance from the early Sun, which led to the freezing out of volatiles at different distances, and remnants from this early epoch are found in the ices of present day comets \citep{2011mumchar, Willacy2015, Dishoeck2020, Raymond2020}. Numerical models predict that most comets formed throughout a trans-Neptunian region \citep{Nesv2017,Vokrouhlick2019} and that gravitational interaction with giant planets moved them into different orbits \citep{Gomes2005, Tsiganis2005, Morbidelli2005, Walsh2011, Levison2011, Batygin2013, brasser2013}. Comets are commonly sub-classified by their dynamical properties. For example, Jupiter Family comets (JFCs) have low-inclination orbits that mostly stay within Jupiter's orbit and likely originated from the Kuiper Belt. Centaurs are low inclination comets in unstable orbits that are in transition between Kuiper Belt Objects (KBOs), also referred to as Trans-Neptunian Objects (TNOs), and JFCs \citep{done15, sarid2019}. Oort Cloud comets (OCCs) have scattered inclinations with orbital distances extending as far as 10,000 - 50,000 au and rarely pass through the inner solar system. The Tisserand parameter with respect to Jupiter, T$_J$, is a numerical value calculated from orbital elements of a comet and Jupiter, and is often used in the classification schema. We adopt the identification of JFCs are comets with 2 \textless T$_J$ \textless 3, while OCCs and Halley type comets (HTCs) have T$_J$ \textless 2 \citep{Levison1996}. HTCs mostly orbit between Saturn and Neptune with higher inclination orbits, and some may originate from the Oort Cloud \citep{Levison1996}. 

Measuring the relative abundances of key diagnostic cometary volatiles is a powerful constraint to solar system formation models \citep{lewis1980, pollack1980, Womack1992Constraints, fegley1999, lodders2003, Ahearn2012, guilbert2015}. The orbital families of comets and the chemical composition of their comae may be interdependent, and through observations we can test models of solar system formation (e.g., \citet{Ahearn2012}). One critical pair of volatiles in comae is CO$_2$ and CO, and their abundance ratio is also useful for constraining physical models of comet nuclei as conveyed by outgassing behavior over different heliocentric distances \citep{prialnik2004, belton09, Mousis2016, sarid2019}. 

The most abundant parent volatile observed in comae is usually H$_2$O, followed by smaller contributions from CO$_2$, CO, and numerous relatively minor species \citep{bockelee2004}. Within heliocentric distance, R$_{Helio}$, of 2.5 au, water-ice sublimation proceeds efficiently and releases other volatiles, icy grains, and dust from the nucleus\citep{Weaver1989, prialnik2004, meech04}. 
Beyond $\sim$ 3 au, water-ice sublimation decreases substantially, and fewer comets show noticeable dust comae \citep{Meech2001, Maz2009, Ivanova2011, Sarneczky2016, Jewitt2017, hui2019, yang2021}. Models of distant activity are generally based on a relatively unprocessed and volatile-rich nucleus and may include significant outgassing of other cosmogonically abundant volatiles in the nucleus, crystallization of amorphous water-ice process, sublimation of icy grains in the coma, jets, landslides, and even impacts \citep{womack17review}. Because of outgassing variations with heliocentric distance, nucleus composition models must incorporate analysis of coma mixing ratios as a function of R$_{Helio}$ \citep{huebnerbenkoff1999}. This paper focuses on CO and CO$_2$ measurements due to their relatively high abundance in comae and low sublimation temperatures, which are expected to contribute to cometary activity \citep{oot2012, Ahearn2012, reach13, bau15}. CO and CO$_2$ production rates are also useful for assessing the C/O ratio in comae, which is a key diagnostic of solar nebula chemistry in the comet-forming region \citep{oberg2011}.


CO can be observed in comae from both ground- and space-based telescopes at a variety of its molecular transitions; however, due to severe telluric contamination and lack of a permanent dipole moment, CO$_2$ can only be observed directly from space. Indirect measurements of CO$_2$ emission are sometimes possible via spectroscopy of related CO Cameron bands and forbidden oxygen emission, but are difficult to carry out and not often derived with simultaneous CO production rates. Thus, the limiting volatile for this mixing ratio is CO$_2$, which has been detected or measured to significant limits in far fewer comets than CO. The largest single survey of simultaneous CO$_2$ and CO measurements thus far was made with the AKARI space telescope \citep{oot2012}. AKARI detected CO$_2$ in seventeen comets, but simultaneous detections or significant upper limits for CO were achieved in only eleven. The CO/CO$_2$ production rate ratios from the AKARI survey showed a great deal of variation among the comets with seven (64\%) being CO$_2$-dominant. This was interpreted as possible evidence for a highly oxidized comet formation environment and/or that CO was exhausted near the surface of the comet nuclei \citep{oot2012}. By itself, the AKARI sample size of the CO/CO$_2$ mixing ratio in eleven comets was too small to test models of the predicted roles that comet families or heliocentric distance may have played.  

In order to explore the CO/CO$_2$ mixing ratio in a much larger sample, as well as the parameters of comet families and heliocentric distances, we compiled a dataset of all contemporaneous CO and CO$_2$ measurements from the literature for 25 comets, and their water production rates when available. From this dataset we analyzed the production rate ratios for CO/CO$_2$, CO/H$_2$O, CO$_2$/H$_2$O, (CO+CO$_2$)/H$_2$O. We also present and discuss the C/O ratio, and the Q$_{CO}$/(surface area) and Q$_{CO_2}$/(surface area) ratios derived using published diameters.

\section{Methods and Analysis}
\label{sec:methods and analysis}

We compiled all known published measurements of CO and CO$_2$ production rates in comets and Centaurs. The subset of objects for which both CO and CO$_2$ were detected or measured to a significant limit is given in Table \ref{tab:scopes}. The vast majority of published production rates were calculated under the assumption that CO and CO$_2$ came from the nucleus, while some indicated that CO emission may also arise from an extended source and contribute another 10 - 50\% to the overall CO coma \citep{dis99, Brooke2003, gunnarson2008, biver2018}.  
For the one paper in this survey that did provide production rates for both nucleus and extended sources of CO \citep{combes1988}, we used only the nucleus value for consistency with the other values. The CO$_2$ production rates we used were all assumed to be coming from the nucleus with no extended source contributions. 

Half of the CO/CO$_2$ mixing ratios in this dataset were obtained using data obtained with the same technique (generally from space-borne infrared spectroscopy). The other half were compiled with different techniques and/or not truly simultaneous, but were close in time. 
Since a variety of ground- space-based techniques were used to collect these data, we briefly summarize them and explain how we handled issues that arose. For reference, the methods used for each comet are summarized in Table \ref{tab:scopes}.

\subsection{Overview of measurements techniques for CO and CO$_2$ measurements}
\label{sec:techniques}

\begin{figure}[h!]
\begin{center}
\includegraphics [width=.75\textwidth] {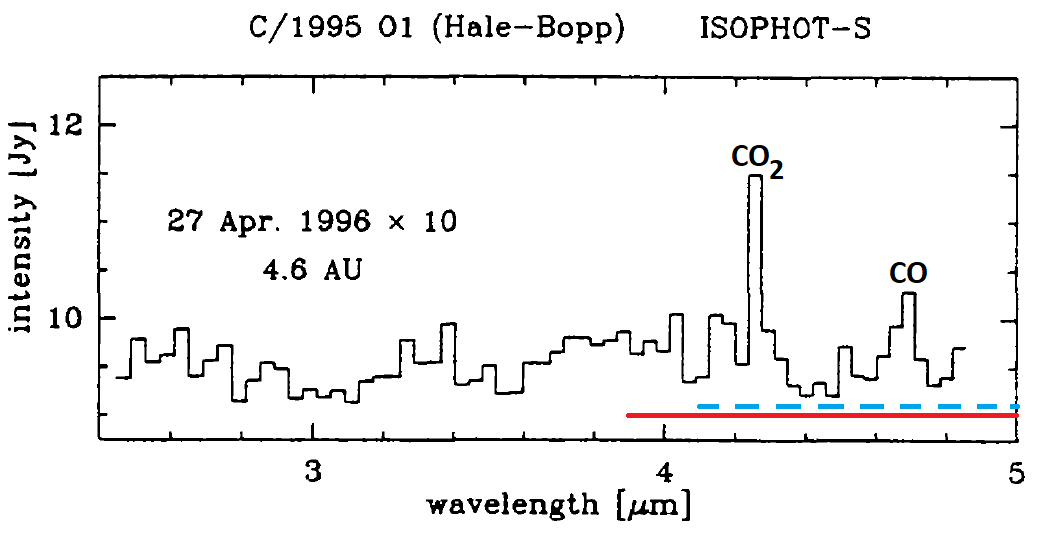}
\caption {\label{fig:FilterSpectraCompare} This figure shows the ambiguity of measuring CO$_2$ (4.26 $\mu$m) and CO (4.67 $\mu$m) production rates with broadband infrared imaging. The wavelength ranges for Spitzer IRAC channel 2 (red solid line) and for NEOWISE channel W2 (blue dashed line) are plotted on an infrared spectrum of comet C/1995 O1 Hale-Bopp obtained with the Infrared Space Observatory (adapted from the Figure 1 of \citet{crovisier1999}). Both CO and CO$_2$ emission bands fall within the channel bandpasses and cannot be distinguished without independent measurements of at least one of the volatiles. This figure also shows that the CO$_2$ emission line is $\sim$ 2x higher than CO's, even though the CO production rate was five times greater, due to differences in fluorescence excitation rates.}
\end{center}
\end{figure}

Most of the Q$_{CO}$/Q$_{CO_2}$ ratios were derived from observations of emission from the anti-symmetric stretch in the $\nu_3$ band of CO$_2$ at 4.26 $\mu$m and the first excited state to the vibrational ground state of CO at 4.67 $\mu$m \citep{crovisier1999, bockelee2004}. Spectroscopic techniques with AKARI, Vega, ISO, and Deep Impact/EPOXI space-based instruments provided simultaneous measurements of these bands in 15 comets (see Figure \ref{fig:FilterSpectraCompare} and Table \ref{tab:observations}). When two aperture sizes were used (such as with the AKARI and Deep Impact instruments) we kept the data that were obtained with the larger aperture since it had the higher signal-to-noise ratio \citep{feaga14}. This may have introduced a small amount of emission from extended sources The production rates are 10-20\% higher in the larger aperture sizes, but correcting for this is beyond the scope of this work and is probably within the uncertainties quoted for these comets. Wide bandpass filters spanning $\sim$ 4 - 5 $\mu$m were also used to measure these CO and CO$_2$ emission bands towards dozens of comets with the Spitzer \citep{reach13} and NEOWISE space telescopes \citep{bau15}. Unfortunately, this method collects photons from both species and cannot be used to derive individual measurements of each molecule (see Figure \ref{fig:FilterSpectraCompare}). It is also impossible to discern whether CO or CO$_2$ is observed from the filtered images without separate spectra of one of the volatiles. Except in rare cases (such as comets known to be either CO$_2$ or CO dominated, ie., 103P or 29P, respectively), the production rates in publications from the Spitzer and NEOWISE surveys are not good approximations for individual molecules, but instead serve as flags for targets of study and may be useful to track overall CO+CO$_2$ production. Due to the inability to determine a CO/CO$_2$ mixing ratio from these data alone, we did not include any production rates that were reported solely from the Spitzer or NEOWISE photometry techniques. We did, however, use these data when independent CO data were also obtained simultaneously, which permitted a CO$_2$ production rate to be inferred (see Section \ref{sec:R2example}). 

Because of a factor of $\sim$ 11.6 in fluorescence efficiencies (see Section \ref{sec:R2example}), CO$_2$ will often have a higher line strength than CO at 4.5 $\mu$m wavelength region, even if the CO production rate is several times higher. An example of this can be seen in Figure \ref{fig:FilterSpectraCompare}, where the calculated CO production rate of Hale-Bopp was 5 times higher than that of CO$_2$, and yet the CO$_2$ line strength is noticeably taller in the spectrum. Moreover, the presence of a CO$_2$ line and absence of a CO emission line does not automatically convey that CO$_2$ is the dominant volatile, due to the significant difference in their fluorescence efficiencies. Numerous examples exist in the AKARI dataset \citep{oot2012} where CO$_2$ was detected and CO was not, and yet CO's production rate upper limit is higher than that derived from the CO$_2$ detection. Thus, caution should be used when interpreting which gas is likely dominating in 4.5 $\mu$m imaging or spectral data.

We included measurements of the CO J=2-1 rotational transition at 230 GHz that were obtained of C/2016 R2 using millimeter-wavelength spectroscopy with the Arizona Radio Observatory Submillimeter Telescope \citep{wie18}. Due to its rotational symmetry, CO$_2$ does not have not any pure rotational transitions. 

For several comets, we inferred the CO/CO$_2$ mixing ratio using indirect measurements. For example, CO$_2$ production rates relative to water can be inferred from optical spectra of forbidden oxygen lines when cometary and telluric emission is significantly separated by Doppler shift \citep{Decock2013,mckay16,Raghuram2014}. These lines are formed when oxygen in an O-bearing molecule photodissociates and provides a green line at 5577\r{A} and two red lines at 6300\r{A} and 6364\r{A}. Photochemical models predict that H$_2$O, CO$_2$, and CO in comae could all be the parents of the dissociation of oxygen, and that if the green to red doublet flux ratio (G/R) is larger than 0.1, then it is likely that either CO or CO$_2$ is the main parent and the percentage relative to H$_2$O can be estimated \citep{Festou1981, cochran2001}. CO is not considered to be a large contributor to most cometary [OI] lines since the dissociation energy of CO to O($^1$D) and O($^1$S) is higher than the ionization energy \citep{raghuram2020}. By itself, this method does not provide a way of discerning whether CO or CO$_2$ is present, and measurements from another technique are needed to see which molecule is dominant. For comets 46P and 21P, we used CO$_2$ production rate values that were derived from OI lines using the Harlan Smith Telescope and Subaru Telescope, respectively, while also using infrared spectroscopy with NASA IRTF iShell to establish the CO production rates \citep{Roth2020,Mckay21, Shinnaka2020}. 

CO$_2$ and CO production rates were also derived from ultraviolet spectra of the CO Cameron and CO 4$^{th}$ Positive bands. The Cameron bands ($a^3$ $\Pi \rightarrow$ $X^1$ $\Sigma ^+$, central wavelength at 198.6 nm) are a spin forbidden transition of a CO molecule that is a dissociation product of CO$_2$, and thus can be used to infer CO$_2$ production rates \citep{raghuram2012}\footnote{There are additional possible contributions from the electron-impact excitation of CO, the electron-impact dissociation of CO$_2$ and the dissociative recombination of CO$_2 ^+$ \citep{feldman1997}.}. This technique works well for CO-depleted comets (comets 103P, C/1979 Y1 Bradfield, C/1989 X1 Austin, C/1990 K1 Levy), or comets which are observed with high enough spectral resolution (comets C/1996 B2 Hyakutake, 9P/Tempel, and 103P/ Hartley 2)\footnote{Comet 103P was a good candidate for using the CO Cameron bands to derive CO$_2$ production rates due to being both a carbon-depleted comet and observed with high spectral resolution.} \citep{bockelee2004}. 
Emission from the CO 4$^{th}$ Positive bands ($A^1$ $\Pi \rightarrow$ $X^1$ $\Sigma ^+$, central wavelength at 158.9 nm) can be modeled to provide CO production rates. The CO Cameron and 4$^{th}$ Positive bands were also sometimes measured simultaneously to provide CO$_2$ and CO production rates, respectively, for comets C/1979 Y1 Bradfield, C/1989 X1 Austin, and C/1990 K1 Levy using the IUE SWP \citep{feldman1997} and for comet C/1996 B2 Hyakutake using the HST FOS \citep{mcphate1999}. Additional observations were used from the CO 4$^{th}$ Positive bands obtained with the HST COS (for comet 103P) and ACS (for comet 9P) instruments \citep{weaver2011,feldman2006}. 

\begin{deluxetable*}{llll}
\renewcommand{\arraystretch}{0.85}
\tabletypesize{\scriptsize}
\tablecaption{Overview of Simultaneous CO and CO$_2$ measurements in 25 comae \label{tab:scopes}}
\tablecolumns{5}
\tablewidth{0pt}
\tablehead{
\colhead{Instrument} &
\colhead{Molecular transitions} &
\colhead{Comets}&
\colhead{References}
}
\startdata
     AKARI IRC/NC & CO$_2$ ($\nu_3$), CO v(1–0) & 22P, 29P, 81P, 88P, 144P &1\\
     " & " & C/2006 OF2, C/2006 Q1 &\\
     " & " & C/2006 W3, C/2007 N3&\\
     " & " & C/2007 Q3, C/2008 Q3 &\\
     Vega IKS & CO$_2$ ($\nu_3$), CO v(1-0) & 1P & 2\\
     ISO ISOPHOT-S & CO$_2$ ($\nu_3$), CO v(1-0) & C/1995 O1 & 7\\
     Deep Impact/EPOXI HRI & CO$_2$ ($\nu_3$), CO v(1-0) & 9P, 103P, C/2009 P1 & 4, 5, 9\\
   Spitzer IRAC & CO$_2$ ($\nu_3$), CO v(1-0) & C/2016 R2, C/2012 S1, 29P & 8, 10, 14, 17, 18\\
   ARO Submm Telescope & CO J=2-1 & C/2016 R2
     & 8, 14\\
     Smith Optical Telescope &[OI] for CO$_2$ & 46P & 11 \\
     Subaru Telescope HDS &[OI] for CO$_2$ & 21P & 20 \\
     IRTF iShell & CO v(1-0) & 21P, 46P & 11, 19\\
    IUE SWP & CO 4$^{th}$ Positive, CO Cameron & C/1979 Y1, C/1989 X1 &13, 15\\
    " & " & C/1990 K1 & \\
    HST FOS & CO 4$^{th}$ Positive, CO Cameron & C/1996 B2 & 16 \\
     HST COS/ACS & CO 4$^{th}$ Positive 
     & 9P, 103P &3, 6\\
     ROSETTA ROSINA & CO$_2$, CO & 67P & 12\\
\enddata
\tablerefs{ 1: \citet{oot2012}, 2: \citet{combes1988}, 3: \citet{weaver2011}, 4: \citet{ahearn2011}, 5: \citet{feaga2007}, 6: \citet{feldman2006}, 7: \citet{crovisier1999}, 8: \citet{mckay2019}, 9: \citet{feaga14},
10: \citet{lisse2013},
11: \citet{Mckay21},
12: \citet{Combi2020},
13: \citet{feldman1997}
14: McKay, A. et al. (in prep),
15: \citet{tozzi1998}, 16: \citet{mcphate1999}, 17: \citet{wier2019}, 18: \citet{Meech13}, 19:\citet{Roth2020}, 20: \citet{Shinnaka2020}
}
\end{deluxetable*}

We also used measurements obtained by spacecraft instruments for five comets. Vega 1's IKS IR spectrometer provided CO and CO$_2$
data for gas production rates in 1P/Halley's coma \citep{combes1988}. Using the HRI-IR spectrometer, Deep Impact obtained measurements of CO$_2$ in 9P/Tempel 1 during its original mission \citep{feaga2007}, and then of CO$_2$ in 103P/Hartley 2, and both CO and CO$_2$ in C/2009 P1 during its extended EPOXI mission \citep{ahearn2011, feaga14}. No detections 
were achieved for CO with the HRI-IR spectrometer for 9P/Tempel 1 and 103P/Hartley 2 \citep{feaga2007,ahearn2011}, so we used measurements of the CO 4$^{th}$ Positive bands that were obtained simultaneously with the CO$_2$ HRI-IR measurements to derive the CO/CO$_2$ mixing ratios \citep{feldman2006, weaver2011}. For 67P, we used analysis of data from the Rosetta ROSINA DFMS mass spectrometer for CO and CO$_2$ from \citet{Combi2020}.

\subsection{Inferring CO$_2$ from CO+CO$_2$ infrared photometry and independent CO measurements}
\label{sec:R2example}

We disambiguated the combined CO+CO$_2$ emission at $\sim$ 4.5 $\mu$m in Spitzer and NEOWISE infrared imaging data for three comets by using simultaneous independently measured CO emission. As an example, we explain the steps used with C/2016 R2 data. First, Spitzer data was used to calculate a gas production rate with the assumption that all the excess emission in the 4.5 $\mu$m range, after dust subtraction, came from CO. This is typically called the ``CO production rate proxy", Q$_{CO_{proxy}}$ \citep{bau15}. The CO production rate was also calculated using independent measurements, Q$_{CO_{independent}}$ (in this case, the Arizona Radio Observatory 10-m Submillimeter Telescope) was subtracted from the Spitzer value, leaving a residual value, Q$_{residual}$:

\begin{equation}
\label{eq:spitzer}
Q_{CO_{residual}} = Q_{CO_{proxy}} - Q_{CO_{independent}}.
\end{equation}

The residual CO proxy production rate was then converted to a CO$_2$ production rate following:

\begin{equation}
\label{eq:gfactor}
Q_{CO_2}=Q_{CO_{residual}}{ {g_{CO}} \over {g_{CO_2}}}, 
\end{equation}

where the fluorescence efficiencies at 1 au are $g_{CO}$ = 2.46 x 10$^{-4}$ sec$^{-1}$ and $g_{CO_2}$ = 2.86 x 10$^{-3}$ sec$^{-1}$ \citep{1983crovenc}.

At heliocentric distances less than $\sim$ 1.8 au, Spitzer CO+CO$_2$ observations can have significant dust contamination \citep{bauer2021}. However, the data we used were obtained from comets well beyond that distance: C/2016 R2 was at 2.76 au and 4.73 au, C/2012 S1 was at 3.34 au, and 29P at 6.1 au. Thus, we do not think that there is any significant thermal dust signal contamination in the Spitzer data. 

We provide additional details for inferring CO$_2$ from C/2012 S1 and 29P in Section \ref{sec:indcomets}. 

\subsection{When CO and CO$_2$ measurements are obtained with different techniques and/or are not simultaneous}
\label{sec:nonsimultaneous}

CO$_2$, CO, and H$_2$O can rarely be detected simultaneously while using the same instrumentation, so some ratios are derived using different techniques and at slightly different times. We acknowledge that this mixing of techniques makes it difficult to quantify uncertainties of these derived ratios. Because modeling of CO, CO$_2$, and H$_2$O in comets is significantly hindered by lack of measurements 
we carefully included some measurements that were obtained with a different techniques. We alleviated this concern by analyzing case studies of when the individual volatiles were obtained by multiple techniques for a comet and checking for agreement in production rates that were obtained at overlapping times.

For example, with comet 9P, measurements of water production rates from the infrared are in excellent agreement with those derived from OH at 3080 Angstroms (Section \ref{subsec:9P}). Similarly, the water production rates are in good agreement for those derived via IR and optical techniques for 46P (Section \ref{subsec:46P}), and for those derived from IR, Radio and UV techniques for C/2016 R2 (Section \ref{subsec:2016R2}). Regarding CO, the production rates for C/2016 R2 derived via IR and mm-wavelengths techniques, including with different telescopes and observing teams, agree very well \citep{mckay2019}. Similar excellent agreement was also found for CO in 29P with IR and mm-wavelength techniques \citep{senay1994, Paganini2013, Wierzchos2020, bockelee2022}. CO$_2$ and H$_2$O production rates that were obtained for Hale-Bopp with a variety of different methods also showed good agreement \citep{weaver97, crovisier1999, biver2002}. In particular, the CO$_2$ production rates for Hale-Bopp that were inferred from the ultraviolet CO Cameron bands are within $\sim$ 15\% of the production rates derived from IR detections of CO$_2$ three days later \citep{weaver97,crovisier1999}. This is discussed in more detail in Section \ref{subsec:1995O1}. This gave us confidence in combining production rates derived from different techniques. In Section \ref{sec:indcomets} we provide more details of the comparisons for the multiple techniques observed for individual comets.

Since comets undergo heating that can significantly alter production rates and sometimes trigger outbursts, it is important to measure CO and CO$_2$ at the same time to ensure that the mixing ratio is not affected by changes in nucleus volatile production. There are many published measurements of CO and CO$_2$ for the same comet that were obtained weeks, months and sometimes years apart that we did not use to calculate a CO/CO$_2$ mixing ratio. There are not many instances where CO and CO$_2$ were measured at the same time or even approximately the same time, and all of the values are listed in Table \ref{tab:observations}, with the exception of 67P, which has a plethora of values of both volatiles from the Rosetta mission. The selected values of Q$_{CO}$ and Q$_{CO_2}$ for 67P in this study are representative of the comet’s behavior at the time and are not outliers. Unfortunately, there are not enough published data for all the other comets to establish long term production rates of CO2, especially in coordination with CO.

There are five comets for which measurements were not simultaneous, but were taken close enough in time to use: C/2016 R2, 9P, 21P, 46P, and 103P. CO and CO$_2$ data were obtained 11 hours apart for R2, 1.7 days apart for 9P, 1.1 days apart for 103P, 3 days apart for 46P, and 7 days apart for 21P. Extensive monitoring of these comets with all techniques indicated that there were no outbursts or significant variation in outgassing rates over the observed time periods, so we are confident about their mixing ratios being stable over these short time periods \citep{feaga2007, feldman2006, ahearn2011, weaver2011, Mckay21, Roth2020, Shinnaka2020}.\footnote{We used the midpoint in time between the measurements for determining the heliocentric distance and true anomaly values in Table \ref{tab:observations}.} 

\subsection{Uncertainties of production rate ratios}

Error bars for the production rate ratios in Figures \ref{fig:indivratios}, \ref{fig:trueanomaly180}, \ref{fig:dynageOCC}, \ref{fig:COH2Oall}, \ref{fig:CO2H2Oall}, and \ref{fig:coco2h2o} were calculated with the standard propagation of error using the production rate uncertainties. Unfortunately, publications do not always include these values. When production rate uncertainties were provided they are given in the tables, i.e., see Table \ref{tab:observations}, for 1P, C/1996 B2, C/2006 W3, C/2008 Q3, and C/2009 P1. These uncertainties ranged from 14\% - 34\% and the average was 20\%. In order to estimate a reasonable uncertainty for the mixing ratios for the comets without published uncertainties, we assumed a 25\% uncertainty for Q$_{CO}$ and Q$_{CO_2}$, which we do not include in the tables. Throughout Section \ref{resultsanddis} we also discuss how distantly active comets are more likely to have CO, CO$_2$, and/or H$_2$O detected, which might be a contributing factor to the lack of JFCs that we see beyond 3.5 au. This could contribute a bias in the results that is not accounted for in the numbers provided in the paper.

\bigskip
\section{Results and Discussion}
\label{resultsanddis}

\subsection{Aggregate findings: CO/CO$_2$}
\label{aggregate}

CO$_2$ and CO production rates were measured in the comae of 25 comets over the heliocentric distance range of 0.71 au to 6.18 au, with values ranging from Q$_{CO_2}$ = (0.06 -- 740) $\times$ 10$^{26}$ molecules sec$^{-1}$ (0.4 -- 5,400 kg sec$^{-1}$) and Q$_{CO}$ = (0.2 -- 3,000) $\times$ 10$^{26}$ molecules sec$^{-1}$ (1 -- 14,000 kg sec$^{-1}$) (see Table \ref{tab:observations}). The production rate ratios Q$_{CO}$/Q$_{CO_2}$ are plotted against heliocentric distance in Figure \ref{fig:indivratios}. In order to facilitate comparison, we identified which volatile was dominant by marking these divisions in Figure \ref{fig:indivratios}, Figure \ref{fig:trueanomaly180}, and Figure \ref{fig:dynageOCC} using these dividing lines:

\begin{itemize}
\item Q$_{CO}$/Q$_{CO_2}$ $\geq$ 4/3 (CO-dominant)

\item 2/3 $<$ Q$_{CO}$/Q$_{CO_2} <$ 4/3 (CO and CO$_2$ comparable)

\item Q$_{CO}$/Q$_{CO_2}$ $\leq$ 2/3 (CO$_2$-dominant)
\end{itemize}

These limits serve as a guide to help explore the volatile dominance shown by each comet. Upper limits to the production rate ratios were only included when they were less than or equal to 2/3. Only one lower limit was used (for Centaur 29P) where CO was detected and only a significant limit was set for CO$_2$ emission. As Figure \ref{fig:indivratios} shows, the ratios vary significantly from Q$_{CO}$/Q$_{CO_2}$ $\sim$ 0.01 for 103P/Hartley 2 at 1 au to Q$_{CO}$/Q$_{CO_2}$ $>$ 98 for 29P at 6 au.

Using the above divisions and looking at all objects as an aggregate, thirteen comets (52\%) fall in the CO$_2$-dominant category, nine (36\%) are CO-dominant, and three (12\%) have comparable amounts of CO and CO$_2$ in their coma (see Table \ref{tab:results}). However, a closer look indicates that the mixing ratio may follow a trend with heliocentric distance, and there may be differences between comet families. We investigate these possibilities in the following sections.

\subsection{Taxonomical families compared against CO/CO$_2$}
\label{cometfamilies}

A comet's family assignment is typically made based on its current orbital parameters and deduced formation history (see Section \ref{sec:intro}). Comae compositional variations may be due to differences in formation scenarios and/or physical processing histories of comet nuclei after formation \citep{Tsiganis2005,obrien2006, mandt2015, Mumma2003,2011mumchar}. In this section we examine whether there are measurable differences between Q$_{CO}$/Q$_{CO_2}$ ratios among the family groups of OCCs and JFCs. In Table \ref{tab:results} each cometary family is quantified based on their CO/CO$_2$ ratios.\footnote{This survey of CO/CO$_2$ ratios has only one HTC (1P/Halley) and one Centaur (29P/Schwassmann-Wachmann). While these two objects are included in the figures, there are not enough objects in each category to discuss these family groups separately for the CO/CO$_2$ ratios.}

\begin{figure}
\begin{center}
\includegraphics [width=0.99\linewidth] {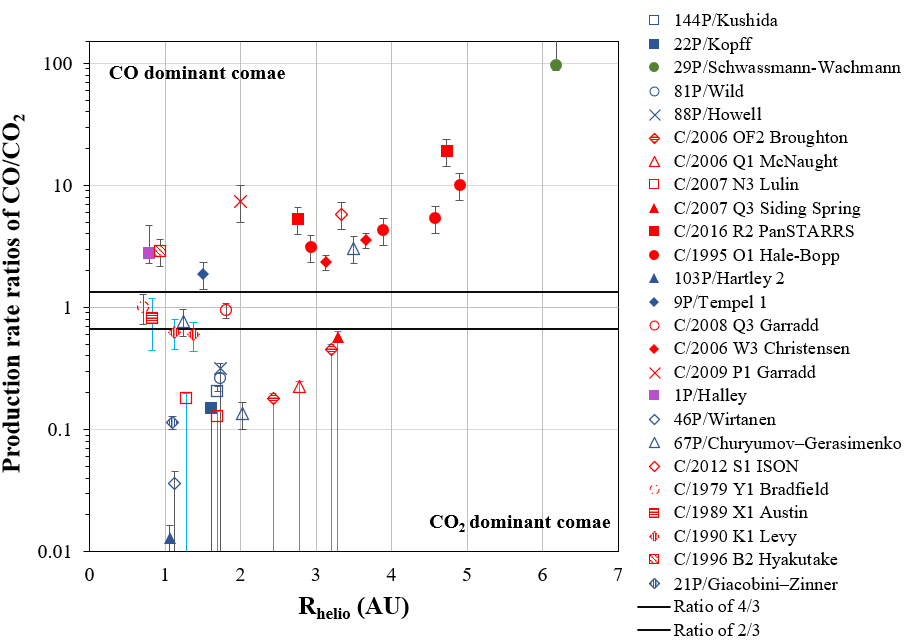}
\caption{ \label{fig:indivratios} Q$_{CO}$/Q$_{CO_2}$ production rate ratios are plotted vs. heliocentric distance for 25 comets. The objects are identified by name and are color coded to indicate family or class (red = Oort Cloud, blue = JFC, purple = Halley Type, green = Centaur). Two horizontal lines are drawn for values at $\pm$ 33\% of 1, and comets within these two lines are considered to have comparable values of CO and CO$_2$ in their coma. Comae with Q$_{CO}$/Q$_{CO_2}$ $\geq$ 4/3 are considered to be CO-dominant and those with Q$_{CO}$/Q$_{CO_2}$ $\leq$ 2/3 are CO$_2$-dominant. Comets that are dynamically new are denoted with blue error error bars. Eight out of the nine JFCs are CO$_2$-dominated. The fourteen OCCs at first appear evenly distributed across the three categories, but including dynamical age divides them into two groups (see Figure \ref{fig:dynageOCC}). The aggregate data are consistent with CO becoming increasingly dominant in most comae beyond 3.5 au, possibly because CO requires a lower sublimation temperature than CO2. However, CO is not necessarily expected to dominate at large distances based on volatility effects alone. Since CO ice sublimates more efficiently than CO2 at large heliocentric distances, it could also be expected that CO should have been exhausted from at least the top layers of OCC nuclei long ago. Even so, mass loss of material during the perihelion passage might open deeper layers, which may be associated with the very high gas and dust production rates typical for many OCCs near perihelion.
}
\end{center}
\end{figure}

As Figure \ref{fig:indivratios} shows, eight of the nine JFCs fall in the CO$_2$-dominant category. 9P is the only JFC that has a CO-dominant coma, with Q$_{CO}$/Q$_{CO_2}$ $\sim$ 2, as previously noted by \citet{Ahearn2012}. Given that most JFCs are thought to originate from Centaurs, it would be useful to see whether the CO/CO$_2$ mixing ratio changes as a comet moves from the Centaur to JFC region. The lone Centaur with a CO/CO$_2$ ratio, 29P, is CO-dominant, while most of the JFCs are CO$_2$-dominant; however, since 29P is at $\sim$ 6 au, it is subjected to much less heating than the JFCs, most of which were observed within $\sim$ 2.5 au, so 29P's high CO abundance may be a solar heating effect. In contrast, the fourteen OCCs are more evenly distributed across the three categories with 36\% of the OCCs (red) are CO$_2$-dominant, 43\% are CO-dominant, and $\sim$ 21\% have comparable amounts in their coma. Thus, looking at all the measurements combined, it seems as though JFCs might be more dominated by CO$_2$ than OCCs. However, there may be a selection effect at work. All of the JFCs were observed within 3.5 au, while the OCCs ratios were obtained across a larger range of heliocentric distances. The lack of data for JFCs beyond 3.5 au may be at least partly due to the fact that JFC nuclei tend to be smaller than those of OCCs. Smaller comets are generally less active at larger heliocentric distances and also fainter. Thus, the values at large heliocentric distances may not be typical of all distant comets (for example, consider the larger nuclei of Hale-Bopp and 29P at 6 au), and additional measurements of comets, especially JFCs, beyond 4 au are needed to better constrain cometary models.

Before exploring this difference in CO/CO$_2$ mixing ratios between JFCs and OCCs further for possible formation or physical processing causes, we search for possible solar heating effects since we have data ranging from 1 to 6 au.

\begin{deluxetable*}{lrccr}
\renewcommand{\arraystretch}{0.85}
\tabletypesize{\scriptsize}
\tablecaption{Overview of Comet Families in CO/CO$_2$ Survey \label{tab:results}}
\tablecolumns{5}
\tablewidth{0pt}
\tablehead{
\colhead{Family Class} &
\colhead{CO$_2$} &
\colhead{CO} &
\colhead{CO and CO$_2$} &  
\colhead{Total}\\
&
\colhead{Dominant} &
\colhead{Dominant} &
\colhead{Comparable} &  
\
}
\startdata
     JFC & 8 & 1 & - & 9\\
     OCC & 5 & 6 & 3 & 14\\
     HTC & - & 1 & - & 1\\
     Centaur & - & 1 & - & 1\\
     Total & 13 & 9 & 3 & 25 \\
\enddata
\end{deluxetable*}

\subsection{CO/CO$_2$ and heliocentric dependence or true anomaly}
\label{sec:heliocentric}

The aggregate data are consistent with an increasing CO/CO$_2$ production rate ratio with increasing distance from the Sun. In addition, all comae beyond $\sim$ 3.5 au are dominated by CO (see Figure \ref{fig:indivratios}). Although there are not many distantly active comets in this sample, they are all consistent with preferential release of CO over CO$_2$ from the nucleus. Coma abundances at large heliocentric distances may be a poor reflection of nucleus abundances, and instead, measurements of comets closer to the Sun might be a better indicator of nuclear abundances. Nonetheless, the larger heliocentric distance data are included to test models of the behavior of the release of volatiles from the nucleus as a function of heliocentric distance \citep{huebnerbenkoff1999}. Although we know that water-ice, the dominant volatile component in most comets, does not sublimate very efficiently at large distances, the outgassing of CO and CO$_2$ at large distances is not as well understood. The CO-dominance observed in comae beyond 3.5 au may be due to a few comets which have higher amounts of CO and are intrinsically more productive. This could also be caused by a heating effect on the nucleus and preferential release of volatiles for all comets. There are not many measurements of CO and CO$_2$ in JFCs observed beyond 3 au and so we cannot rule out a selection effect. More simultaneous measurements of CO and CO$_2$ production rates are needed in JFCs beyond 3 au to provide strong observational constraints.

\begin{figure}[h!]
\begin{center}
\includegraphics [width=0.95\textwidth] {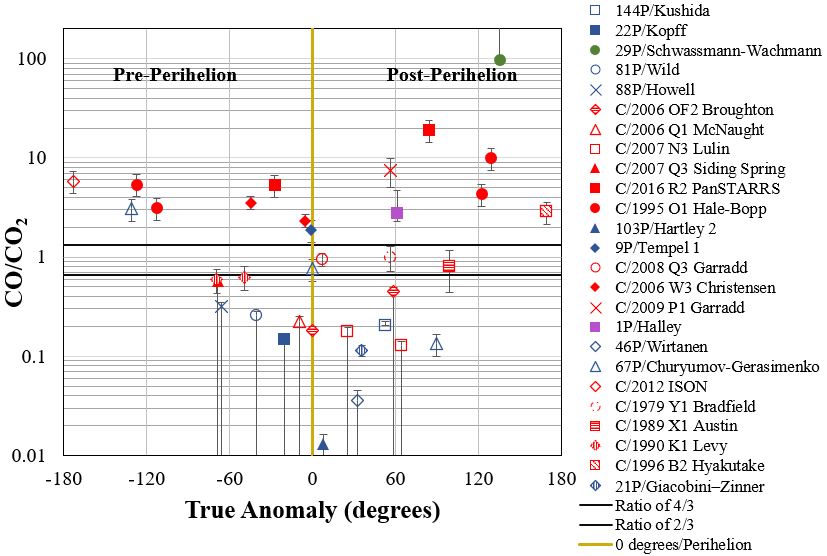}
\caption {\label{fig:trueanomaly180} The CO/CO$_2$ production rate ratio is plotted against the true anomaly, which highlights the difference between pre- and post-perihelion values. Perihelion is at 0 degrees and is denoted with a yellow vertical line. Two horizontal lines are drawn within $\pm$ 33\% of 1 and the color codes are the same as with Figure \ref{fig:indivratios}. The post-perihelion production rate ratios show a slight increase in CO post-perihelion for comets C/2016 R2 and C/1995 O1 Hale-Bopp. Comets with true anomalies greater than $\pm$ 120 degrees (both incoming and outgoing) are CO-dominant.
}
\end{center}
\end{figure}

Another way to explore possible heating effects is to examine the mixing ratio as a function of true anomaly, which is the angle between the periapsis and observed position relative to the Sun, and is often used to look for changes in a comet's behavior as it goes through perihelion passage. In Figure \ref{fig:trueanomaly180} we do not see a significant difference between the CO/CO$_2$ mixing ratio in pre- and post-perihelion of any comets either as an aggregate group, or based on dynamical classes. However, all comets observed beyond $\pm$ 120 degrees are CO-dominant, which is consistent with also being beyond 3.5 au.

We have an opportunity to check for heliocentric dependence in this mixing ratio for four comets: C/2006 W3, C/1995 O1 Hale-Bopp, C/2016 R2, and 67P. Comets C/1995 O1 Hale-Bopp and C/2016 R2 were always CO-dominant and became slightly more CO-dominant after perihelion. C/2006 W3 produced slightly more CO than CO$_2$ when it was farther from the Sun pre-perihelion, but it was not measured post-perihelion. The Q$_{CO}$/Q$_{CO_2}$ ratio of 67P varied in a complicated way during its perihelion passage. 67P was first measured as CO-dominated at $\sim$ 3.5 au pre-perihelion, then switched to CO$_2$-dominated from 3-2.5 au, and then produced approximately equal amounts of CO and CO$_2$ from 2 au to perihelion at 1.2 au. Thereafter, it switched to a CO$_2$-dominated coma, which it maintained for the most of the post-perihelion passage\footnote{At 3.5 au post-perihelion comet 67P had a CO/CO$_2$ = 0.12, showing the opposite behavior of the 3.5 au pre-perihelion detection switching from CO to CO$_2$ dominance behavior}. Since most of the 67P measurements showed it to have far more CO$_2$ than CO, we put it in the CO$_2$ dominant category, although there are many many (20+) observations for the CO and CO$_2$ we only include three representative values for 67P in the figures. Clearly, the production of CO and CO$_2$ in 67P's coma over time is complex and may be affected by the bi-lobed structure, which sometimes produced different amounts of volatiles \citep{Combi2020} (see Section \ref{subsec:67P} for further discussion).

29P is a unique case since it is a Centaur and has a near-circular orbit at $\sim$ 6 au and is thus difficult to include in a heliocentric distance analysis. CO$_2$ was only measured once (as an upper limit) in 29P (and calculated indirectly twice, see Section \ref{subsec:29P}, so it is not possible to study the mixing ratio's change over time. However, the visible dust coma and CO production rate both show signs of a heliocentric dependence for the quiescent outgassing behavior, as well as a reported correlation of outbursts seen post-perihelion, see \citet{Wierzchos2020} for details.

Due to the small number of comets observed, the significantly increased production of CO over CO$_2$ in distant comets can only be considered a preliminary result. If this holds for other comets, then it may be an important clue to the mechanical release of volatiles in distantly active comets and Centaurs when water-ice sublimation is less active. It is not clear how much CO-dominance beyond 3.5 au is due to a selection effect of comets with higher amounts of CO being intrinsically more productive and how much is due to a heating effect on the nucleus and the release of volatiles since CO requires a lower temperature to sublimate than CO$_2$.

Therefore, when considering comae beyond 3.5 au, and in the absence of information on both species, it may be appropriate to assume the dominant volatile is CO rather than CO$_2$, as originally noted in Figure 9 in \citet{bau15}. For example, when analyzing gaseous emission in NEOWISE and Spitzer CO+CO$_2$ images for comets beyond $\sim$ 3.5 au, when independent data is not available for either molecules, the default volatile proxy should be CO. In some cases, NEOWISE and Spitzer gas production rates values for comets beyond 3.5 au are provided in the literature as CO$_2$ values. Converting them to CO is straightforward and would increase the gas production rates by a factor of $\sim$ 11.6, the ratio of the fluorescence efficiencies (see Section \ref{sec:R2example}). In addition, many comets within 3.5 au may produce a significant fraction of CO, so caution should be observed when relying on published CO$_2$ production rates from Spitzer and NEOWISE studies.

\subsection{Dynamical Age (1/$a_0$) of Oort Cloud Comets}
\label{sec:dynamicalage}

The number of perihelion passages a comet has made is directly related to the total heating its nucleus has experienced, which may in turn affect its coma composition. The inverse of the original semi-major axis, 1/$a_0$, is often used as a parameter for exploring solar heating in OCCs, since these comets have sustained stable orbits for a long period of time and are much farther from the Sun than JFCs or HTCs \citep{ahearn1995}. Furthermore, if we look at OCCs by their so-called dynamical age, we might expect that dynamically new (DN)\footnote{Comets that are considered to be making their first trip from the Oort Cloud are referred to as ``dynamically new'' and are typically identified by 1/$a_0$ \textless 10$^{-4}$ au$^{-1}$ \citep{oort51,dyb2001, Horner2003,dones2004}.} comets would have experienced less processing from the Sun than dynamically older (DO) ones. Consequently, DN comet nuclei are typically assumed to have retained their primordial composition which should be evident in their comae near the Sun, including the CO/CO$_2$ mixing ratio.


\begin{figure}[h!]
\begin{center}
\includegraphics [width=0.95\textwidth] {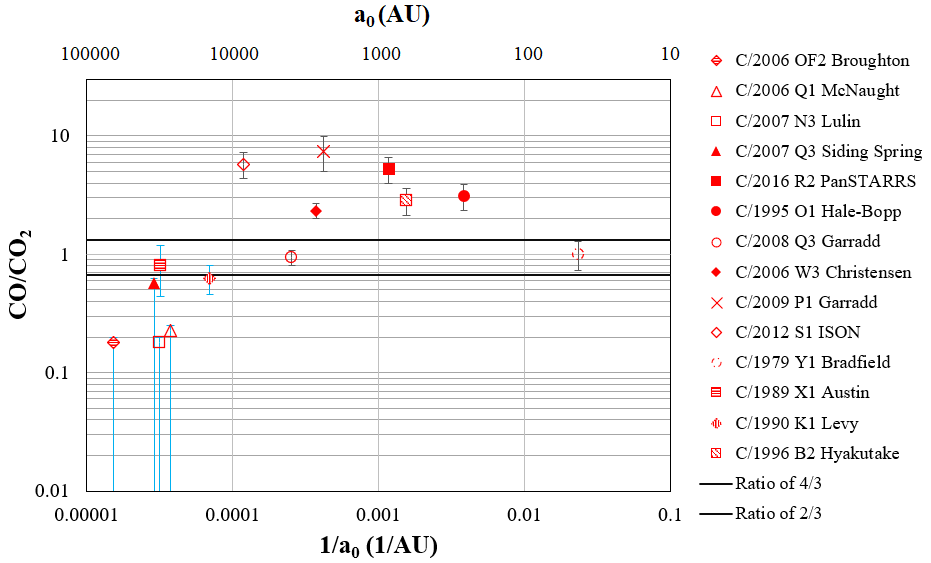}
\caption {\label{fig:dynageOCC} Dynamical age, as probed with the inverse of the original semi-major axis, 1/a$_0$, is plotted for fourteen Oort Cloud comets within 3.5 au. We also plot the original semi-major axis, a$_0$, scale across the top axis. Dynamically new (DN) comets are often defined as those with 1/a$_0$ $<$ 10$^{-4}$ ${au}^{-1}$, which applies to the six left-most comets (C/2006 OF2 Broughton, C/2006 Q1 McNaught, C/2007 N3 Lulin, C/2007 Q3 Siding Springs, C/1989 X1 Austin, C/1990 K1 Levy). DN comets have the lowest CO/CO$_2$ ratios in this sample and a trend may be present with increasing CO produced with more thermal processing from the Sun (moving toward the right). OCCs on their first trip to the inner solar system produce more CO$_2$ than CO when within 3.5 au of the Sun, while those on their second or later trip tend to be CO-dominant. This is interesting, since CO is much more volatile than CO$_2$, and might be expected to have outgassed more than CO$_2$ over billions of years. Instead, this can be explained if CO ice is depleted from the outermost layer of the nucleus by galactic cosmic radiation during Oort Cloud storage, and then sublimates more readily after the outside cosmic-ray processed layer is eroded during its perihelion passage by the Sun, which exposes deeper layers where CO ice is still present \citep{maggiolo2020}. As with previous figures, two horizontal lines are drawn within 33\% of 1, and comets within these lines are considered to have comparable values of CO and CO$_2$ in their coma. DN comets are are highlighted by having blue error bars rather than black error bars.
}
\end{center}
\end{figure}

In Figure \ref{fig:dynageOCC} we plotted the CO/CO$_2$ production rate ratios 
of each OCC observed within 3.5 au as a function of 1/a$_0$\footnote{These 1/a$_0$ values are from the MPC database and are listed in Table \ref{tab:observations}.}. Within 3.5 au, water-ice sublimation proceeds efficiently and we can assume a quasi-steady-state coma abundance for CO and CO$_2$. We use 3.5 au as the cutoff rather than an even more conservative 2.5 au, because the identity of a CO-dominated or CO$_2$-dominated coma of these comets did not change between 2.5 and 3.5 au, and using 3.5 au provides more data points to probe whether CO or CO$_2$ was more dominant in the coma. For example, all of the CO/CO$_2$ ratios that we have for C/2006 OF2 shows that it had more CO$_2$ than CO in its coma, whether we use the measurements at 2.4 au or 3.2 au.\footnote{Using 2.5 au as the cutoff, instead, does not change any conclusions in this section.} We include only one data point for each comet and use the lowest heliocentric distance measurement available in order to minimize possible solar heating effects.

\begin{deluxetable*}{lrccccc}
\renewcommand{\arraystretch}{0.85}
\tabletypesize{\scriptsize}
\tablecaption{Abundances Relative to Water for Comets under 2.5 au \label{tab:RelWater}}
\tablecolumns{7}
\tablewidth{0pt}
\tablehead{
\colhead{Volatile/s} &
\colhead{\# JFCs} &
\colhead{\# OCCs} &
\colhead{\# HTCs} & 
\colhead{\# Interstellar} & 
\colhead{Total \# Comets} &  
\colhead{\% of Water (Median)}
}
\startdata
     CO & 6 & 23 & 3 & 1 & 32 & 3  \textpm{ 1}\\
     CO$_2$ & 14 & 8 & 1 & - & 23 & 12  \textpm{ 2}\\
     CO+CO$_2$ & 9 & 8 & 1 & - & 18 & 18   \textpm{ 4}\\
\enddata
\end{deluxetable*}

Interestingly, Figure \ref{fig:dynageOCC} shows that comets on their first trip to the inner solar system produce more CO$_2$ than CO, while those on their second or later trips tend to produce more CO. This is the opposite of what models typically predict, which is that comets with the longest time in the Oort Cloud should have the more pristine composition. This result is consistent with what was found in a smaller study of only five comets \citep{ahearn1995}, which at the time was attributed to small-number statistics and/or by selection effects in the data set. Now with almost triple the number of comets in the 2012 study we see the same trend that DN comets produce more CO$_2$ relative to CO than the other Oort Cloud comets that have made multiple passes through the inner solar system. 

This result may be explained by a model in which galactic cosmic ray impacts damaged ices in the outer layers of OCCs (while JFCs and Centaurs were relatively protected in the ecliptic). This led to a preferential loss of CO over CO$_2$ in these top layers down to several meters (due to CO's much lower sublimation temperature and the modest energy increase from the cosmic rays), which then is observed in the low CO/CO$_2$ mixing ratios during the comets' first approach to the Sun. This highly processed layer is significantly eroded and more well-preserved ices are exposed and outgas during subsequent perihelia \citep{gronoff2020, maggiolo2020}. Thus, the comae of DO comets are more pristine than those of DN comets. Figure \ref{fig:dynageOCC} also shows that there may be a trend of increasing CO produced with more solar passages (increasing 1/a$_0$) for dynamically older comets, which should be confirmed with more measurements. Further efforts to confirm this observational result and to test models of comet nucleus processing from both solar heating and cosmic irradiation over very long periods are strongly encouraged.


One exception to the trend of increasing Q$_{CO}$/Q$_{CO_2}$ with increasing 1/a$_0$ is C/1979 Y1, which is the dynamically oldest comet in this study and yet has comparable amounts of CO and CO$_2$, instead of producing far more CO. Including HTC 1P/Halley in Figure \ref{fig:dynageOCC}, would put it all the way to the right as the most solar radiation processed comet in the figure, just beyond C/1979 Y1, and it has a CO/CO$_2$ mixing ratio of $\sim$ 3. Interestingly, C/1979 Y1 may more accurately belong in the HTC category based on its orbital characteristics (Section \ref{sec:indcomets}). Thus, 1P's CO/CO$_2$ mixing ratio more closely resembles that of the other DO OCCs than C/1979 Y1. Comets Y1 and 1P might be another part of the trend that link DO comets and HTCs, 
or that 1P and/or C/1979 Bradfield are outliers. Adding all of the available data (including the JFCs) to Figure \ref{fig:dynageOCC} would introduce more CO$_2$-dominant comae to the far right side. It is interesting that the CO/CO$_2$ mixing ratios of the dynamically new OCCs are comparable to those of the highly-processed JFCs and yet are possibly explained by different types of processing - cosmic irradiation for the DN and solar irradiation for the JFCs. 

Whether this result that DN comets produce more CO$_2$ than CO at perihelion holds for all Oort Cloud comets can be tested this year with the DN comet C/2017 K2 (PanSTARRS), which first became active at large distances. CO was detected in K2 at 6.7 au with Q$_{CO}$ = 1.6 x 10$^{27}$ molecules/sec \citep{yang2021}. K2 has 1/a$_0$ = 4.4x10$^{-5}$ (according to MPC) and thus would be located between C/2006 Q1 and C/1990 K1 on the figure. If K2's nucleus has preferentially lost CO relative to CO$_2$ in its outer layers due to cosmic ray processing, then it is likely to have Q$_{CO}$/Q$_{CO_2}$ \textless 2/3 and will produce at least as much CO$_2$ as CO when it is near perihelion on Dec 19, 2022 at 1.8 au. Future measurements of both CO and CO$_2$ emission in C/2017 K2, and other DN comets, are needed to confirm the unexpected results in Figure \ref{fig:dynageOCC} and cosmic ray processing model of \citet{gronoff2020, maggiolo2020}.

\begin{sidewaysfigure}
\centering
\includegraphics [width=0.99\textwidth] {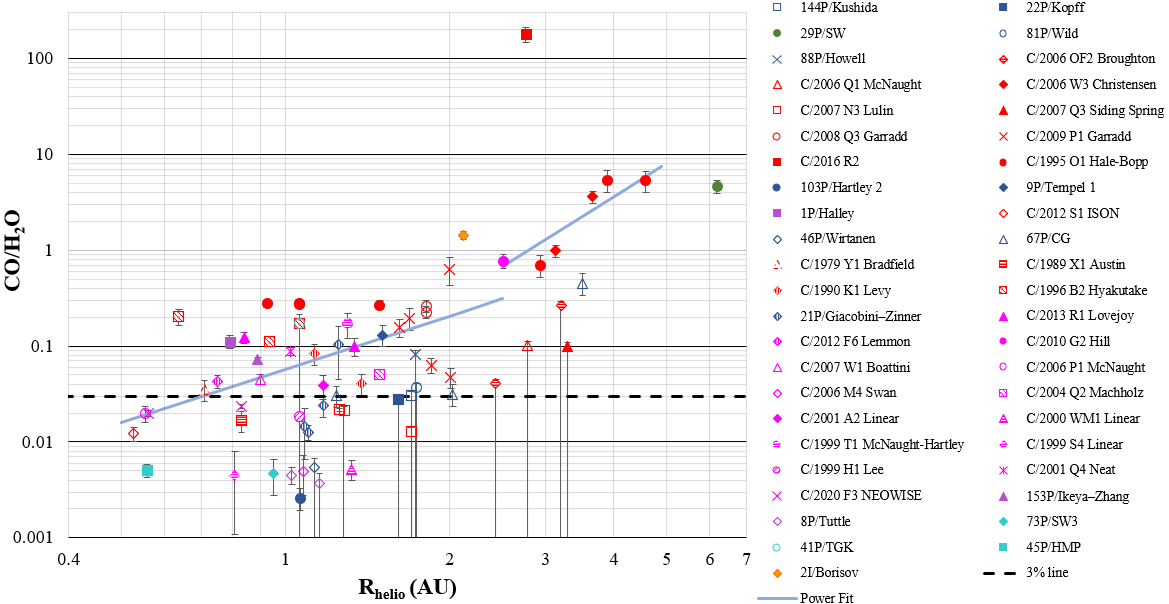}
\caption {\label{fig:COH2Oall} CO/H$_2$O data from 45 comae. Comets from the previous figures are plotted using the same color scheme. Comets that did not have simultaneous CO and CO$_2$ detections are plotted with new colors: added OCCs are magenta, added JFCs are cyan, and one added interstellar comet is orange (2I/Borisov). The dashed horizontal line shows the median of all data (except Borisov) obtained within R$_{Helio} < 2.5$ au: (Q$_{CO}$/Q$_{H_2O}$)$_{median}$ = 3.3 $\pm$ 1.3\%, with a higher Q$_{CO}$/Q$_{H_2O}$ value for most OCCs compared with JFCs. This plot also shows how anomalous C/2016 R2's composition was compared to all other comets and is consistent with a very distant formation region beyond the CO snow line, while most other comets have low CO/H$_2$O ratios consistent with formation within the CO snow line. Outside 2.5 au, the fractional abundance of CO to H$_2$O steadily increases, which is likely due to the decreased efficiency of water-ice sublimation at larger heliocentric distances. Overall, a two-slope fit to this data results in Q$_{CO}$/Q$_{H_2O}$ $\sim$ R$_{Helio}^{1.6}$ between 0.5 au to 2.5 au, and Q$_{CO}$/Q$_{H_2O}$ $\sim$ R$_{Helio}^{3.6}$ from 2.5 au to 4.6 au.
The fits were calculated using detections only and did not include upper limits, nor were they weighted by uncertainties. Beyond 2.5 au, the slope for CO/H$_2$O is noticeably steeper than that for CO$_2$'s/H$_2$O (see Figure \ref{fig:CO2H2Oall}). Accounting for the upper limits would further increase the CO/H$_2$O slope for the large R$_{Helio}$ comets, but also increase the scatter in the fit. At larger distances there are fewer measurements, and the fit from 2.5 au to 4.6 au should be considered with caution. The data are listed in Table \ref{tab:observations} and Table \ref{tab:supobservations}.
}
\end{sidewaysfigure}

\subsection{CO/H$_2$O, CO$_2$/H$_2$O, (CO+CO$_2$)/H$_2$O, and C/O production rate ratios}
\label{subsec:COCO$_2$H$_2$O}

\begin{figure}[h!]
\begin{center}
\includegraphics [width=0.99\textwidth] {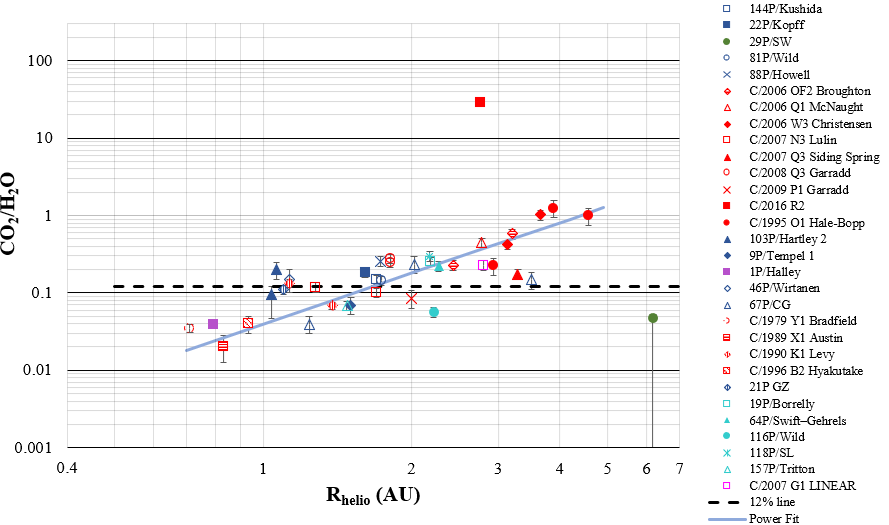}
\caption {\label{fig:CO2H2Oall} The fractional production rate of CO$_2$ to water is plotted against heliocentric distance and appears to follow a relationship of Q$_{CO_2}$/Q$_{H_2O}$ $\sim$ R$_{Helio}^{2.2}$ from 0.71 to 4.58 au. In contrast to CO, the CO$_2$ to water ratio increases less steeply with heliocentric distance and shows a tighter correlation with heliocentric distance (compare to Figure \ref{fig:COH2Oall}). The dashed line indicates the median value of Q$_{CO_2}$/Q$_{H_2O}$ = 12\% $\pm$ 2\% for all comae below 2.5 au. The data are provided in Table \ref{tab:observations} and Table \ref{tab:supobservations}. 
} 
\end{center}
\end{figure}

\begin{figure}[h!]
\begin{center}
\includegraphics [width=0.99\textwidth]{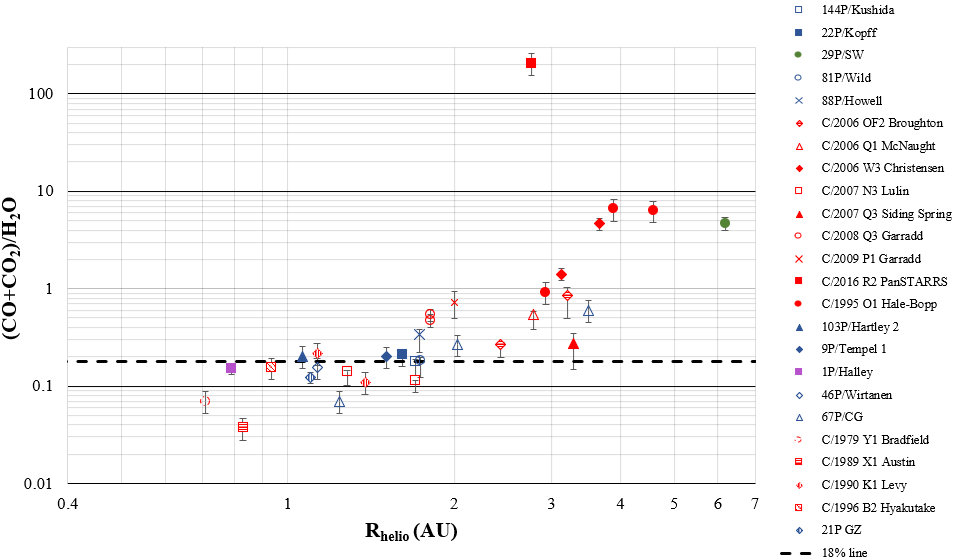}
\caption {\label{fig:coco2h2o} This figure plots the ratio of the total inorganic carbon-bearing volatile (as represented by CO and CO$_2$) to water production rates as a function of heliocentric distance. The same color conventions are used for the different comet families as the previous figures. The dashed line shows the location of the median production rate ratio of Q$_{CO+CO_2}$/Q$_{H_2O}$ = 18 $\pm$ 4\% for all comets within 2.5 au. 
}
\end{center}
\end{figure}

We also used the compiled production rates to examine the relative amounts of CO and CO$_2$ with respect to H$_2$O and to explore the total carbon volatile budget in comae. CO, CO$_2$, and H$_2$O production rates are listed in Table \ref{tab:observations} when all three volatiles were measured contemporaneously, and in Table \ref{tab:supobservations} when either of the paired production rates of CO and H$_2$O, or CO$_2$ and H$_2$O, were obtained at the same time. Production rate ratios of Q$_{CO}$/Q$_{H_2O}$ and Q$_{CO_2}$/Q$_{H_2O}$ as a function of heliocentric distance are plotted in Figure \ref{fig:COH2Oall} and Figure \ref{fig:CO2H2Oall}, respectively. We also examined the combined CO and CO$_2$ production rates with respect to water in Figure \ref{fig:coco2h2o} and the estimated volatile carbon/oxygen ratios in Figure \ref{fig:c/o}. We discuss implications of these figures in this section.

Figure \ref{fig:COH2Oall} shows the CO/H$_2$O production rate ratios plotted for 45 comets as a function of heliocentric distance and coded for orbital family. The CO/H$_2$O ratio for comae within 2.5 au (which includes 32 of the 45 comets) ranges from 0.3\% to 26\% with a median of 3.3 \textpm{ 1.3\%}.\footnote{The uncertainty is the standard error of the median.} Within 2.5 au, the OCCs tend to have a higher CO/H$_2$O ratio than the JFCs, which may be in part due to different formation environments. This result is based on values from six JFCs, three HTCs, and 23 OCCs and should be confirmed with more observations. The median was calculated using detections in both Table \ref{tab:observations} and Table \ref{tab:supobservations}, and no upper limits were used. If instead of a median, we calculate an average, we find CO/H$_2$O = 6.4 \textpm{ 1.3\%} within 2.5 au, which is consistent with what was reported for the average of 24 comets of 5.2 \textpm{ 1.3\%} in another study \citep{dello2016}. 

The production rate ratio of Q$_{CO}$/Q$_{H_2O}$ for all comae over the entire range of heliocentric distances observed (Figure \ref{fig:COH2Oall}) is noticeably steeper than that of Q$_{CO_2}$/Q$_{H_2O}$ (Figure \ref{fig:CO2H2Oall}). Moreover, the CO fractional abundance may exhibit a change in slope at $\sim$ 2.5 au, being flatter interior and steeper further out. To illustrate this point and for easier comparison with the CO$_2$ data, we calculated two slopes using a least squares fit\footnote{The fits did not include upper limits, the outlier C/2016 R2, nor the interstellar comet Borisov, and the fits were not weighted by the uncertainties.}, resulting in Q$_{CO}$/Q$_{H_2O}$ $\sim$ 
R$_{Helio}^{1.6}$ between 0.5 au to 2.5 au, and Q$_{CO}$/Q$_{H_2O}$ $\sim$  
R$_{Helio}^{3.6}$ from 2.5 au to 4.6 au. The CO/H$_2$O data have a significant amount of scatter and we caution against over-interpreting these values, but it is worth noting that this possible change in slope for CO/H$_2$O at 2.5 au coincides with the distance within which the water ice sublimation starts to become less efficient. When looking at the fits to Figure \ref{fig:COH2Oall} the large heliocentric measurements of comets are biased towards comets that are active and productive enough to be measured, and so tend to favor larger comets. 
Since the upper limits were not included in the fits to Figure \ref{fig:COH2Oall} a possible bias might be introduced again against lower CO comets being included in the analysis. If upper limits were included, the slopes would be even steeper and the scatter would be increased.

The CO$_2$/H$_2$O production rate ratios were compiled for 23 coma measurements within 2.5 au, and they range from 2\% to 30\% with a median value of (Q$_{CO_2}$/Q$_{H_2O})_{median}$ = 12 \textpm{ 2\%}. For comparison, the median is $\sim$ 30\% smaller than the 17\% value derived by \citet{oot2012}. We attribute this difference to our sample size being almost twice as large as the AKARI survey, and to the fact that most of these additional comets had much lower fractional CO$_2$/H$_2$O abundances. 

In contrast to CO, the CO$_2$ to water ratio increases less steeply with heliocentric distance and shows a tighter correlation with water from 0.7 to 4.6 au with Q$_{CO_2}$/Q$_{H2O}$ $\sim$ R$_{Helio}^{2.2}$ (again excluding C/2016 R2). This may mean that production of CO$_2$ is intimately tied to water production in a way that CO is not. In contrast, the CO production rate has a much stronger response to solar heating than CO$_2$, which is consistent with it being a much more volatile ice. Interestingly, despite apparent possible strong ties to water, the aggregate CO$_2$/H$_2$O data do not show evidence for a break in slope at $\sim$ 2.5 au where water-ice sublimation typically changes significantly for comets.

\begin{figure}[h!]
\begin{center}
\includegraphics [width=0.5\textwidth] {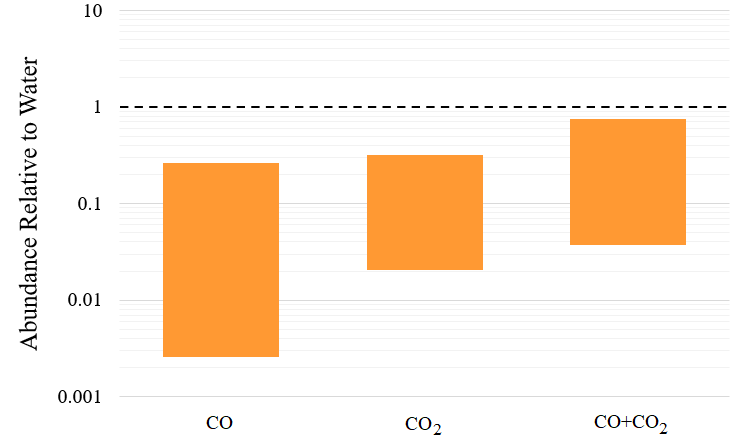}
\caption {\label{fig:relrates} The ranges of fractional production rates of CO, CO$_2$, and CO+CO$_2$ with respect to water H$_2$O for comae within 2.5 au. The CO/H$_2$O ratio has a wider range than CO$_2$, which could be partly due to the fact that there are almost twice as many CO detections than CO$_2$. Although the (CO$_2$+CO)H$_2$O abundance ratio range in this figure is large, measurements for most comets fall within a much narrower range of 18\% $\pm$ 4\%, as shown in Figure \ref{fig:coco2h2o}.
}
\end{center}
\end{figure} 

We also examined the total CO+CO$_2$ production rates as a proxy for the total inorganic volatile carbon budget, since the vast majority of carbon will be found in these two volatiles \citep{ahearn1995}. Only 18 of the 25 comets that have CO and CO$_2$ production rates were observed below 2.5 au. We restricted our calculation to comets observed below 2.5 au, which should minimize possible bias introduced by differential outgassing of CO and CO$_2$ that we see farther out. The median production rate ratios of (Q$_{CO}$ + Q$_{CO_2}$)/Q$_{H_2O}$ for all comets within 2.5 au is 18 \textpm{ 4\%}, although we point out that there is some substantial deviation from this value for a few comets (notably C/1989 X1 Austin with $\sim$ 3.7\% and C/2009 P1 Garradd at $\sim$ 70\%). However, most of the comets are consistent with the median value of 18\%, which indicates that the total amounts of CO and CO$_2$ produced within 2.5 au may be conserved for most comets and is close to the $\sim$ 20\% value proposed by \citet{Ahearn2012, 2021Lisse}. Thus, the (Q$_{CO}$ + Q$_{CO_2}$)/Q$_{H_2O}$ ratio for comets observed within 2.5 au could serve as a possible cosmogonic indicator providing insight on formation temperatures of the early solar system \citep{faggi2018, lippi2020}. This 
could support the theories that comets retain a strong amount of their natal composition. We are cautious because his study is based on a relatively small sample size, and more observations would be helpful to see how pervasive this is. We can also see that the CO+CO$_2$ seems to be well conserved in comets, and is indicative of comets forming in the same region. This is from the fact that when we look at comets observed within 2.5 au in Figure \ref{fig:coco2h2o} (where there are about nine JFCs and eight OCCs), and most of the comets, regardless of their dynamical family, stay relatively level/flat along the median abundance ratio (i.e. no strong outliers like C/2016 R2) is suggestive that comets share a similar natal formation region.

Next, we consider the carbon/oxygen ratio tied up in gas comae as a probe of the cometary formation environment (c.f., \citet{oberg2011, eistrup2018}). We use CO, and CO$_2$ to represent carbon, since they are the largest carbon-contributors by far to the gas coma, and we use H$_2$O, CO, and CO$_2$ for the main contributors of oxygen (see \citet{milam2006,remijan2008, dello2016}). In Figure \ref{fig:c/o} we plot the carbon to oxygen ratio that we calculated using

\begin{equation}
C/O = { {Q_{CO}+Q_{CO_2}} \over{Q_{CO}+2Q_{CO_2}+Q_{H_2O}}}
\end{equation}

\noindent for all comets where all three species were detected at the same time (Table \ref{tab:observations}). A caveat is that the total cometary C/O ratio will also include the solid dust/grain component, but here we address only the volatile sources of carbon and oxygen as a probe of the gas and ice components.

The heating effects of solar radiation can be seen in Figure \ref{fig:c/o} where comets that are closer to the Sun produce significantly more water compared to carbon monoxide and carbon dioxide (resulting in a lower overall ratio). Within $\sim$ 2.5 au, where water sublimation is vigorous and individual comae probably achieved approximate compositional stability, we find C/O$_{average}$ $\sim$ 15\% and C/O$_{median}$ $\sim$ 13\%, which are similar to values found by \citet{Seligman2022} for a sample of solar system comets, and is consistent with these comets forming within the CO snow line (i.e., C/O $<$ 0.2) \citep{Seligman2022}. 
We see no distinction between the JFCs and OCCs with respect to C/O ratios within 2.5 au. This may be explained by models which predict that both Kuiper Belt and Oort Cloud reservoirs have similar origins and that they both formed within the region of the giant planets (i.e. 5-30 au), and were later moved to their present positions (see discussion by \citet{dones2015}). 
The low C/O gas-phase values are consistent with forming within the CO snow line \citet{oberg2011}). Comets with higher values, such as C/2016 R2 with C/O $\sim$ 0.9, may have formed much farther out, even beyond the CO snow line. Caution should be used for high C/O values derived for comets beyond $\sim$ 3.5 au, such as C/2006 W3 Christensen, Hale-Bopp, and 29P, which are likely increased due to the lowered sublimation of water-ice at these distances and these values should not be used to constrain formation models.

\begin{figure}
\begin{center}
\includegraphics [width=0.99\textwidth]{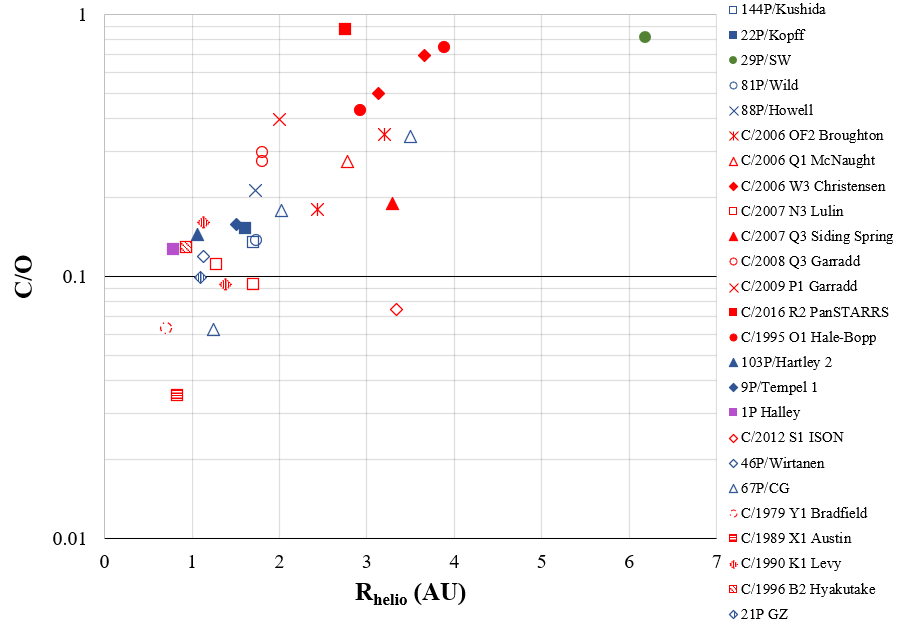}
\caption {\label{fig:c/o} The carbon/oxygen ratio for the gaseous component (determined from CO, CO$_2$ and H$_2$O production rates) is plotted as a function over heliocentric distance. The same color convention is used for the different comet families as the previous figures. There is no discernible difference between the gaseous C/O ratios in JFCs and OCCs, which is consistent with the comet families forming in overlapping regions. The median C/O ratio in the gaseous component for comae within 2.5 au is $\sim$ 13\%, and this low value is consistent with most comets forming within the CO snow line. C/2016 R2's high value of C/O $\sim$ 0.9 is better fit by having formed much farther out, possibly beyond the CO snow line. 
}
\end{center}
\end{figure}

It is possible that water production rates may include substantial contributions from the sublimation of icy grains that have been dragged out from the nucleus, in addition to sublimating directly from water-ice on the nucleus (cf. \citet{ahearn2011, SunshineFeaga2021}). For comets referred to as ``hyperactive'' (such as 21P, 41P, 45P, 46P, and 103P), the amount of water vapor may be substantial and even rival what is produced by direct sublimation of water ice from the nucleus and lead to what is reported as a lower CO/H$_2$O and CO$_2$/H$_2$O production rate ratio than what is actually coming from the comet. Thus for hyperactive comets we might be observing lower fractional abundances of CO and CO$_2$ due to the``over-production'' of water. It is also possible that CO$_2$ ice grains can be ejected from CO which is more volatile. Other ejected ice grains that are from less volatile molecules (sublimate at higher temperatures) being released from gases that are more volatile (sublimate at lower temperatures) would increase the observed production rate of the volatile with the higher sublimation temperature. This would tend to increase the observed production rate of CO$_2$ over CO \citep{SunshineFeaga2021}. No attempt was made to separate those contributions in this analysis and we just use total water production rates. Caution is warranted for interpreting results for fractional water abundances in comets beyond $\sim$ 3 au since the coma abundances at larger heliocentric distances might be a poor reflection of nucleus abundances. The larger heliocentric observed comet data are provided to test models of their volatility as a function of heliocentric distance.

Opacity is another factor that could affect the production rates, especially for the inner coma and very productive comets. Most reported production rates were calculated assuming optically thin conditions and therefore had minimal-to-no corrections applied \citep{feldman2006, feaga2007, oot2012, feaga14, mckay2015}. In other cases, the optically thick inner coma region was small compared to the field of view, and thus it was assumed not to contribute much to the observed flux \citep{oot2012, wie18}. However, if there was a larger optical depth, then this would lower the reported production rate of each respective volatile \citep{beth2019}.

While extended sources of CO have been observed in some comets like 1P, C/1995 O1, C/2016 R2 \citep{eberhardt1987,disanti1999b,Cordiner2022}, instruments like HST, with high spatial resolution, are better equipped to get native source sublimation. In this paper we note that extended sources of CO and possibly CO$_2$ could contribute to the production rates but not to the extent that would dramatically change the results. For example, photodissociation of CO$_2$ and H$_2$CO could contribute a small amount (less than 10\%) to the total CO of 103P \citep{weaver2011}. In another example, there is no evidence for an increase in production rates with larger aperture sizes with Spitzer measurements of C/2016 R2 \citep{mckay2019}.

\subsection{Carbon-depleted comets}
\label{sec:carbon-depleted}

\begin{figure}[h!]
\begin{center}
\includegraphics [width=0.7\textwidth] {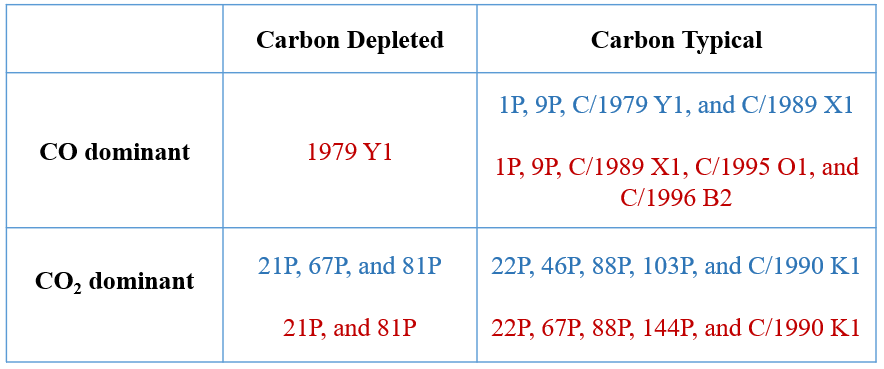}
\caption {\label{fig:carboncontent} The classification of carbon richness (depleted/typical) is compared against the CO and CO$_2$-dominance from this paper. The blue text in the box uses the categorization of depletion or typical carbon content from \citet{ahearn1995}, while the red is the categorization from \citet{cochran2012}.}
\end{center}
\end{figure}

Some of the comets in this CO/CO$_2$ survey have also been classified as ``carbon-depleted'' by their carbon chain abundance in other papers \citep{ahearn1995,cochran2012}, largely based on their C$_2$/CN and C$_3$/CN production rate ratios. 
In \citet{ahearn1995} comets 1P, 9P, 22P, 46P, 88P, 103P, C/1979 Y1, C/1989 X1, and C/1990 K1 are considered typical and comets 21P, 67P, and 81P are considered carbon-depleted. \citet{cochran2012} uses stricter limits for classifying comets, and favors the following as typical: 1P, 9P, 22P, 67P, 88P, 144P, C/1989 X1, C/1990 K1, C/1995 O1, and C/1996 B2; and the following as carbon-depleted: 21P, 81P, and 1979 Y1. Both papers agree that 1P, 9P, 22P, 81P, 88P, C/1989 X1, and C/1990 K1 are typical and comets 21P and 81P are carbon-depleted. These classifications are summarized in Figure \ref{fig:carboncontent}. Looking at the comets that the two teams agree on, carbon-depleted comets 21P and 81P are CO$_2$-dominant in our survey, but so are typical comets 22P, 88P, C/1990 K1. There are also other typical comets, 1P, 9P, and C/1989 X1, that produce more CO than CO$_2$. Thus, we do not see any correlation between carbon-depletion designation and the relative CO/CO$_2$ ratio.

\begin{figure}[h!]
\begin{center}
\includegraphics [width=0.99\textwidth] {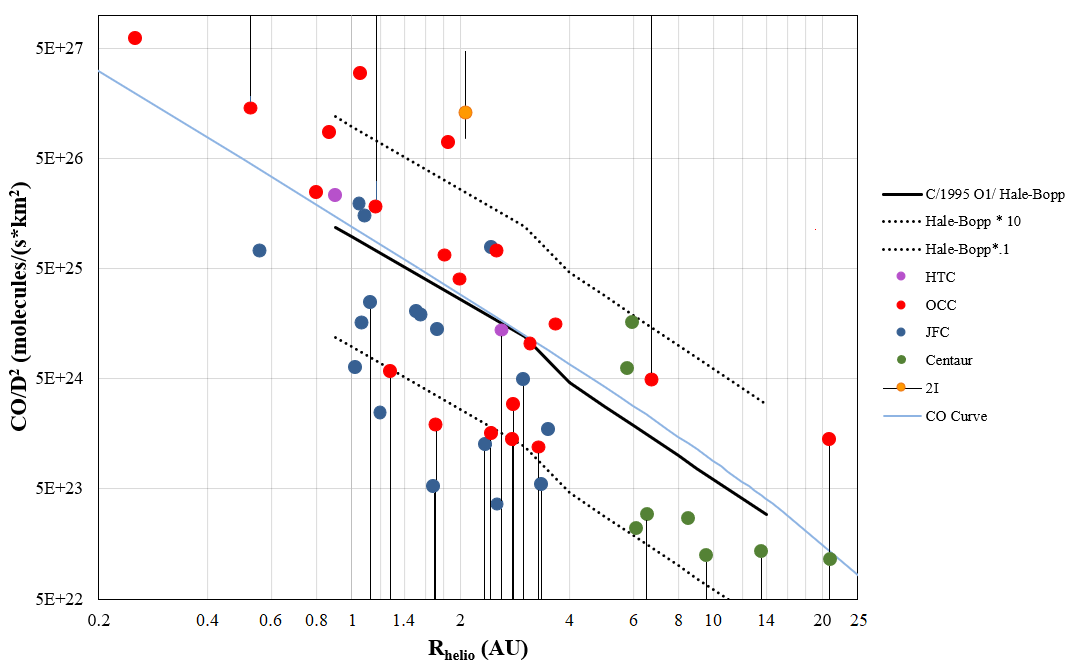}
\caption {\label{fig:subcurve1}Here we plot specific production rate per unit of area, Q$_{CO}$/D$^2$, vs. heliocentric distance. Overlaid in a blue line is the sublimation curve from \citet{sekanina92} for CO for a comet with a diameter of 32 km. When observed over 1 to 2 au, OCCs produce significantly more CO per surface area than most JFCs. One Centaur (29P) shows elevated productivity for its size at 6 au, while all other Centaurs measured to date produce very little CO for their size, or none at all. This could be a response to decreased solar heating of the Centaurs at larger distances. C/2017 K2's CO detection at $\sim$ 6 au and diameter upper limits indicate that the this OCC is producing high amounts of CO for its size, possibly out-producing Hale-Bopp and 29P at this distance. For comparison, the specific production rate for C/1995 O1 Hale-Bopp is shown in solid black line. 
}
\end{center}
\end{figure}

\begin{figure}[h!]
\begin{center}
\includegraphics [width=0.99\textwidth] {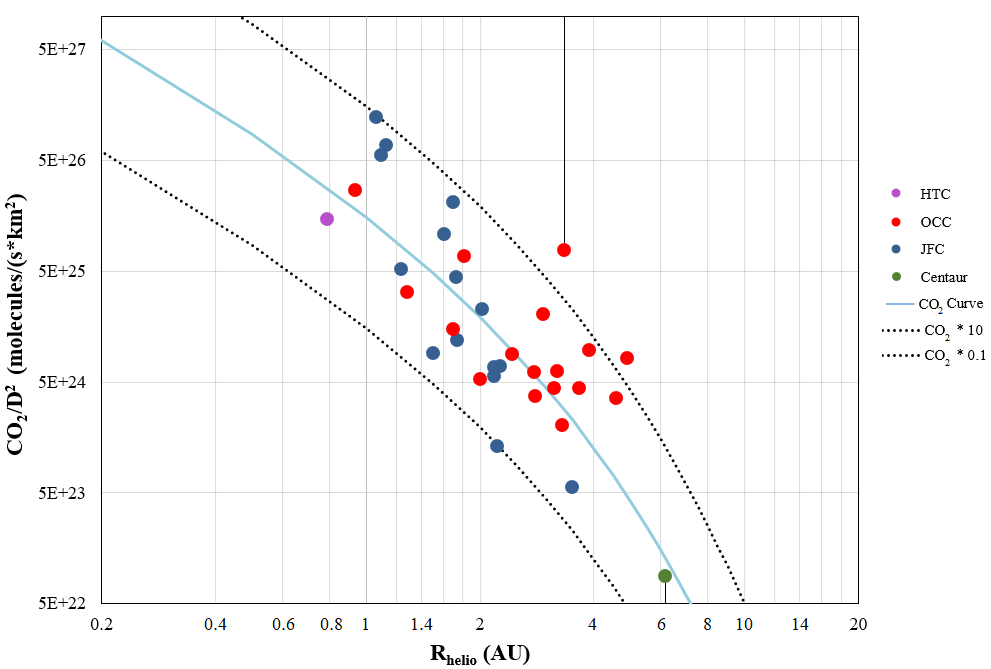}
\caption {\label{fig:subcurve2RS}Here we plot specific production rate per unit of area, Q$_{CO_2}$/D$^2$, vs. heliocentric distance. Overlaid in a blue line is the sublimation curve from \citet{sekanina92} for $CO_2$ for a comet with a diameter of 10 km. The dotted lines above and below the CO$_2$ curve are for the cases when there is ten times more CO$_2$ production and a tenth of the CO$_2$ production, respectively. 
}
\end{center}
\end{figure}

\subsection{Active fraction areas for CO and CO$_2$}
\label{subsec:COD2}


We normalized CO production rates for surface area (assuming a spherical shape) for comets using published diameters and plotted them against heliocentric distance (Figure \ref{fig:subcurve1}). The references for the diameters and the CO productions are in Table \ref{tab:surfobsd}. Three of our comets have significant upper limits for diameters: C/2001 A2, C/2012 S1, and C/2017 K2.

In Figure \ref{fig:subcurve1}, interstellar comet 2I Borisov produces more than 10 times the CO production of C/1995 Hale-Bopp for its surface area and is similar to some other OCCs from our solar system around the distance it was observed. Distantly active comet C/2017 K2 is also in the graph where it shows significant activity at large distances, even above the activity level of C/1995 O1 Hale-Bopp, despite its size not being determined yet. If its nucleus is even smaller than the upper limits used, then C/2017 K2's will have an even higher specific CO production rate for its surface area. 

A CO vaporization curve is superimposed in Figure \ref{fig:subcurve1}, which is the sublimation rate as a function of heliocentric distance for a comet with a 32 km diameter. Interestingly, when looking at 1-4 au in Figure \ref{fig:subcurve1}, the CO specific production rates of most JFCs are an order of magnitude lower than C/1995 O1 Hale-Bopp. In contrast, within 1 au, the CO specific production rates for all OCCs are located above the C/1995 O1 line, and at distances above 1 au most OCCs show less activity. 
The two Halley type comets are similar to the C/1995 O1 Hale-Bopp line, which would support the idea that some HTCs originated from the Oort Cloud \citep{Levison1996}. Centaurs are dispersed with a couple of values from 29P (the lowest from quiescent activity and the highest from outburst activity), much lower detections from Centaurs Echeclus and Chiron, and several upper limits for other Centaurs which are significantly lower than C/1995 O1 Hale-Bopp. The lower CO/D$^2$ values for the Centaurs could either be from incorporation of less CO in the nucleus, or inhibition of CO outgassing at these large distances \citep{Wierzchos2017}. 

Figure \ref{fig:subcurve2RS} shows the normalized CO$_2$ production rates for surface area (assuming a spherical shape) with their respective references listed in Table \ref{tab:surfobsdCO2}. Comet C/2012 S1 is the only comet in this figure with an upper limit diameter. Figure \ref{fig:subcurve2RS} has CO$_2$ vaporization curve for a comet with a 10 km diameter superimposed on the graph. Unlike Figure \ref{fig:subcurve1}, in Figure \ref{fig:subcurve2RS} both JFCs and OCCs follow the behavior of the CO$_2$ sublimation curve with less scatter. There is only one data point in the Q$_{CO_2}$/D$^2$ graph representing HTCs and Centaurs so we can not say anything definitive about either of those classes of objects, other than both appear to fall very near the sublimation line.

\section{Individual comets}
\label{sec:indcomets}
Next we provide additional background information, context, and any additional analysis steps we carried out for the comets analyzed for contemporaneous CO and CO$_2$ measurements. Comets 144P, C/1989 X1, C/1990 K1, C/2006 Q1, C/2007 Q3, and C/2008 Q3 do not have individual subsections. The telescopes used to obtain the CO and CO$_2$ of the individual comets here are listed in Table \ref{tab:scopes}.

\subsection{1P/Halley}
\label{subsec:1P}

We used CO and CO$_2$ production rates obtained from the Vega spacecraft infrared spectroscopy measurements \citep{combes1988}. Production rates were also inferred from ultraviolet spectra of CO with similar numbers, but we instead use infrared spectroscopy because of the direct simultaneous measurements of CO and CO$_2$. 1P/Halley is CO-dominant compared to CO$_2$, which might be consistent with Halley retaining much of its natal abundances.


\subsection{9P/Tempel 1}
\label{subsec:9P}

9P is a notably potato-shaped comet visited by Deep Impact and Stardust spacecraft. It has a short orbital period of six years and a 6 km diameter, with circular features on its surface consistent with a layered nucleus \citep{Thomas2007}. We use the pre-impact measurements of CO and CO$_2$ measured by Deep Impact and HST. The Deep Impact HRI-IR spectrometer detected CO$_2$ and H$_2$O emission on July 4, ten minutes before impact, but did not detect CO \citep{feaga2007}. The HST ACS Solar Blind Channel detected CO 4$^{th}$ Positive gaseous emission a day earlier, which we used for CO production rate \citep{feldman2006}.

For comparison, the pre-impact H$_2$O production rate of 5 $\times$ 10$^{27}$ molecules sec$^{-1}$ derived from OH photometry also on July 4 \citet{schleicher2007} is in good agreement with the 4.6 $\times$ 10$^{27}$ molecules sec$^{-1}$ derived by \citet{feaga2007} on the same day that we used.

The CO/CO$_2$ ratio is 1.88 at R$_{Helio}$ = 1.51 au (pre-impact), and it is the only JFC to be have a CO-dominated coma instead of CO$_2$.

Interestingly, post-impact IR measurements \citep{mumma2005} show a lower CO/H$_2$O ratio than the pre-impact values. It is not clear what causes these differences (different techniques or pre-to-post impact changes), but if the IR values are accurate, and CO$_2$ production relative to water remained constant, then it is possible that 9P's coma may not always be CO dominated compared to CO$_2$.

\subsection{21P/Giacobini-Zinner}
\label{subsec:21P}

This well-studied comet has been observed over many apparitions by several spacecraft, ICE, Pioneer Venus 1, and many ground-based telescopes. It is also considered by many to be the prototypical carbon-depleted comet \citep{moulane2020,Roth2020}.

We calculated the CO/CO$_2$ mixing ratio using a combination of the individual CO and H$_2$O production rates derived from IR data \citep{Roth2020} and a CO$_2$/H$_2$O production rate ratio that was inferred from analysis of the green to red doublet ratio (G/R) of forbidden oxygen that was obtained on October 3, 2018 \citep{Shinnaka2020}. 
We then calculated the CO/CO$_2$ mixing ratio using the CO$_2$/H$_2$O ratio, taken on October 3, 2018 \citep{Shinnaka2020}, and the H$_2$O and CO detections observed seven days later on October 10, 2018 \citep{Roth2020}, according to: 

\begin{equation}
{{Q_{CO}} \over{Q_{CO_2}}} ={{{Q_{CO}} \over{Q_{H_2O}}} \times {{Q_{H_2O}} \over{Q_{CO_2}}}}
\end{equation}

The Q$_{CO}$/Q$_{CO_2}$ ratio = 0.114, which makes it a CO$_2$-dominant comet, like 8 out of 9 JFCs in our sample. 

\subsection{22P/Kopff}
\label{subsec:22P}

22P is a highly active JFC \citep{lamy2002} and a dynamical study indicates that it recently transferred from the Centaur to JFC region, possibly within the last $\sim$ 100 years \citep{Pozuelos2014}. The CO/CO$_2$ ratio is an upper limit of 0.15 \citep{oot2012}, which makes it a CO$_2$-dominant comet, like 8 out of 9 JFCs in our study.

\subsection{29P/Schwassmann-Wachmann}
\label{subsec:29P}

29P was previously considered by many to be a JFC, however orbital dynamical studies showed that 29P is a transitional object between Centaurs and JFCs, and is the first known inhabitant of the ``Gateway'' transition region just beyond Jupiter's orbital distance from the Sun \citep{sarid2019}. 29P is also unusual in that it has a nearly circular orbit at $\sim$ 6 AU, and with an ever-present dust coma, likely driven by CO outgassing \citep{senay1994}. In addition to its constant, low-variable, quiescent activity, it has has several substantial outbursts per year \citep{trigo2008, womack17review, Wierzchos2020} and a persistent 24 $\mu$m ``wing'' feature that is largely created by solar radiation pressure on micron-sized grains \citep{schambeau2015}.

CO is readily measured at millimeter- and infrared-wavelengths, but CO$_2$ emission has not been detected yet in 29P. \citet{oot2012} reported strong upper limits obtained of its 4.67 $\mu$m band with AKARI which yields an average production rate ratio for CO/CO$_2$ $>$ 84, the highest value in our dataset. 

In order to further quantify the CO$_2$ contribution to the coma, we infer the CO$_2$ production rate following the method described in Section \ref{sec:R2example}. To carry out this analysis, we used the total CO+CO$_2$ emission recorded by Spitzer \citep{reach13} and NEOWISE \citep{bau15} infrared space telescopes, CO mm-wavelength data obtained with the IRAM 30-m \citep{bockelee2022}, and visible magnitude data \citep{trigo2010}, all of which were observed during the first half of 2010.

The independent CO data were not obtained simultaneously with the Spitzer data, however, independent visual monitoring showed no evidence of major dust outbursts within a few days of the NEOWISE data \citep{trigo2010}. This Centaur is continuously active and exhibits a relatively constant amount of CO emission and dust, except when in outbursting mode when it increases significantly. A recent study showed that outbursts of CO gas and dust in 29P do not always correlate in time \citet{Wierzchos2020}, so caution is warranted.

The IRAM 30-m observations did not showed gas production rates that varied little over their five observing dates from 2010 February 12 to May 29 with an average of Q$_{CO}$ = 4.6 x 10$^{28}$ mol s$^{-1}$. The visual lightcurve data show no evidence of a dust outburst at the time the Spitzer data were obtained on 2010 Jan 25, so we use the average CO production rate from IRAM for Q$_{independent}$ in Equation \ref{eq:spitzer}, and 11.6 for the ratio of the fluorescence efficiencies (Equation \ref{eq:gfactor}) to derive Q$_{CO_2}$ = (5$\pm$3)x10$^{26}$ mol s$^{-1}$ and a mixing ratio of CO/CO$_2$ $\sim$ 92 using the Spitzer and IRAM data. Within the uncertainties, this is consistent with the AKARI spectroscopic measurements of Q$_{CO_2}$ $<$ 3.5x10$^{26}$ molecules s$^{-1}$ and CO/CO$_2$ $>$ 84 obtained on 2009 Nov 19 \citep{oot2012}. The combined Spitzer and IRAM analysis confirm that CO$_2$ emission has a very small presence in 29P's coma, approximately 1\% of CO and 14\% of H$_2$O\footnote{using the water and production rate is from Herschel measurements in 2010 May \citep{bockelee2022}}. 

The same process was applied to the NEOWISE data, which was also obtained during spring 2010 near when CO was also observed with IRAM. This leads to a Q$_{CO_2}$ $\sim$ 3x10$^{27}$ mol s$^{-1}$. This is $\sim$ eight times lower than the AKARI lower limit in Nov 2009 and the value we derive from Spitzer in Jan 2010, and the mixing ratio is CO/CO$_2$ $\sim$ 16. Although more CO$_2$ is inferred from the NEOWISE data, they are still consistent with CO$_2$ emission being a minor contributor to 29P's coma.

29P clearly has the most CO-dominated coma compared to CO$_2$ in this study. Further studies may clarify how much of this difference is due to intrinsic chemical composition of its nucleus, and how much can be attributed to its relatively far distance from the Sun, which might favor CO release over the less volatile CO$_2$ (see Section \ref{aggregate}). A recent study proposes that the preferential release of CO over CO$_2$ in 29P can be explained largely by amorphous water ice converting to the crystalline state upon increased heating in its relatively new orbit (Lisse et al. 2022, under review).

\subsection{46P/Wirtanen}
\label{subsec:46P}

CO$_2$ emission was not directly detected for 46P, but its relative abundance to water was inferred through analysis of forbidden oxygen (see Section \ref{sec:techniques}), as was a measurement of the water production rate. Three days later, a H$_2$O production rate was also measured with infrared spectroscopy, along with an upper limit to the CO emission. The water production rates obtained on both days (through [OI] and infrared techniques) agreed within 3.6\% and was used to extract a CO$_2$ production rate. We assume that the production rates for CO and CO$_2$ were also stable over these three days (cf. \citet{Mckay21}.) 

\subsection{67P/Churyumov–Gerasimenko}
\label{subsec:67P}

We used simultaneous mass spectroscopy measurements for CO and CO$_2$ from \citep{Combi2020}. Over time, the CO/CO$_2$ production rate ratio spans all three regions (CO-dominant, CO$_2$-dominant, and equal parts). The start of Rosetta's observations of 67P recorded higher CO than CO$_2$. Shortly after, while 67P was still traveling inward, the CO and CO$_2$ production rates were almost comparable with the CO$_2$ production detected being slightly greater than the CO production. After reaching perihelion the CO$_2$ production was greater than the CO production recorded \citep{Combi2020}. The CO and CO$_2$ presented in the figures and table are representative of the production rates at the observed times, and not outliers. Although numerous measurements are available throughout its perihelion passage, we use just three values to represent CO/CO$_2$ behavior of 67P at 1.2 au, 2.0 au, and 3.5 au.

This comet is also known for its rubber duck-like shape, thought to have been the result of two cometesimals colliding. It is sometimes considered as two objects since the interior layers of the lobes are oriented in different directions \citep{Massironi2015}. 67P was measured to be alternately CO and CO$_2$-dominated during its inward journey, most of this variation was determined to be a seasonal effect and partially due to compositional differences between the two lobes of the nucleus \citep{fougere2016,hassig15, Combi2020,Fornasier2021}. Overall it was considered to be a CO$_2$-dominant coma where the average CO/CO$_2$ production rate ratio in the coma was 0.67 \citep{Combi2020}. Interestingly, the smaller lobe has a CO$_2$ spot that is active while the waist, or neck, of 67P has shown more H$_2$O activity and not as much CO$_2$. The changes in the activity of the volatiles are attributed to seasonal effects and heliocentric distance \citep{fougere2016,Combi2020}.

\subsection{81P/Wild 2}
\label{subsec:81P}

In 1974, 81P passed within 0.006 au of Jupiter which dramatically changed its orbit, and simulations indicate that it was previously in a Centaur-like orbit, and that in $\sim$ 8,000 years, 81P may no longer be a JFC \citep{krolikowska2006}. Thus, despite its current JFC designation, 81P may be relatively unprocessed compared to many other JFCs. It has Q$_{CO}$/Q$_{CO_2}$ $<$ 0.26 \citep{oot2012}, which puts it in the CO$_2$-dominant category, similar to 8 out 9 JFCs observed for CO and CO$_2$.

\subsection{88P/Howell}
\label{subsec:88P}

Although measured with multiple techniques, the most constraining measurement of the relative amounts of CO and CO$_2$ in 88P's coma is an upper limit of Q$_{CO}$/Q$_{CO_2}$ $<$ 0.31 derived from AKARI space telescope measurements at 1.73 au on July 3, 2009 based on a detection of Q$_{CO_2}$ = $8.65 \times 10^{26}$ molecules/sec and an upper limit for CO \citep{oot2012}. Two prominent jets were observed on 88P at 1.47 au a month later on Aug 8 with Spitzer IRAC, and CO$_2$ was attributed as the main driver of activity for the jets \citep{reach13}.

\subsection{103P/Hartley 2}
\label{subsec:103P}

Peanut-shaped 103P was visited by the Deep Impact spacecraft, which found it had hyperactive behavior, meaning that it released larger amounts of H$_2$O than expected for the nucleus' surface area \citep{ahearn2011}. Other known hyperactive comets include 21P and 46P. 

We used the CO$_2$ detection obtained with the HRI-IR spectrometer \citep{ahearn2011} and the CO measured from the 4$^{th}$ Positive Bands with HST \citep{weaver2011}. There was an upper limit obtained for the CO$_2$ \citep{weaver2011} (\textless 2 x 10$^{27}$ molecules sec$^-1$) and CO was not detected with the HRI-IR \citep{feaga14}. The upper limit from \citet{weaver2011} is in agreement with the detection from \citet{ahearn2011}. 

The CO/CO$_2$ production rate ratio is 0.01, which makes it the most CO$_2$-dominated coma in our study. CO$_2$ is the primary driver of activity in 103P and likely contributed to dragging out chunks of nearly pure water-ice, which then sublimated in the coma to provide a large fraction of the total H$_2$O gaseous output of the comet \citep{Ahearn2012}.

\subsection{C/1979 Y1 Bradfield}
\label{subsec:1979y1}

C/1979 Y1 was the first comet for which UV measurements were taken over a range of heliocentric distances (0.71 au to 1.55 au) \citep{lev1988}.\footnote{There were two comet Bradfields in 1979, and some results from this time period for this Bradfield were published with the ``1979 X'' designation.} The UV measurements we used are from \citet{feldman1997}, where the CO was obtained via the 4$^{th}$ Positive band, and the CO$_2$ production rate was calculated from CO Cameron Band emission. In Figure \ref{fig:dynageOCC} C/1979 Y1 Bradfield has a smaller semi-major axis than other OCCs in the figure, and with a CO/CO$_2$ = 1.00 it does not follow the behavior of other OCCs that have had multiple trips to thine inner solar system and have CO-dominant comae. It is possible that the low CO/CO$_2$ C/1979 Y1 could be due to the inhibition of CO by a chemical or morphological feature on the comet, or that what we see is the result of the small sample size of OCCs.

The CO/CO$_2$ mixing ratio is 1.00, consistent with comparable amounts of CO and CO$_2$ in the coma. Upon discovery, it was considered to be one of the gassiest comets, comparable to 2P/Encke, with a gas to dust ratio of 22 at 0.8 au \citep{ahearn1981}. Although categorized as coming from the Oort Cloud, its orbital characteristics may be consistent with a Halley type comets \citep{ahearn1995}. For comparison, comet Halley had a ratio of $\sim$ 3.

\subsection{C/1995 O1 Hale-Bopp}
\label{subsec:1995O1}

C/1995 O1 Hale-Bopp was well-studied due to its long-term bright behavior starting upon discovery in 1995. Notable structural features observed in C/1995 O1 Hale-Bopp include three distinct tails: gas, dust, and sodium \citep{Cremonese_1997}, jets, and curved features in the dust coma \citep{warell1999, braunstein1997, womack2021}.
The CO/CO$_2$ ratios are observed for multiple dates at large heliocentric distances and all of the ratios show that C/1995 O1 Hale-Bopp had far more CO than CO$_2$ in its coma. All measurements of CO$_2$ were made when the comet was beyond 3 au. There was CO$_2$ obtained from CO Cameron bands with the HST FOS on September 23, 1996 around the same time as one of the observations from ISO (September 26, 1996). The CO$_2$ from HST was 8.8 x 10$^{28}$ and the from ISO was 7.4 x 10$^{28}$ \citep{crovisier1999}. 
We also see more data, from Table \ref{tab:supobservations} and in Figure \ref{fig:COH2Oall}, showing that CO/H$_2$O was seen in higher amounts compared to most other comets at the same distance, especially at lower heliocentric distances.

\subsection{C/1996 B2 Hyakutake}
\label{subsec:1996B2}

C/1996 B2 Hyakutake is among the brightest comets in the last century and it was discovered about six months after C/1995 O1 Hale-Bopp. C/1996 B2 was the first comet where ethane was detected and in abundance \citep{mumma1996}.

The CO/CO$_2$ ratio is 2.83 and was measured via CO 4$^{th}$ Positive band and confirmed with CO Cameron band measurements \citep{mcphate1999}. Given the high CO abundance, the CO Cameron bands were attributed to CO instead of CO$_2$.

\subsection{C/2006 OF2 Broughton}
\label{subsec:2006OF2}

Comet C/2006 OF2 Broughton was identified as a carbon-chain depleted comet based on its CN and C$_3$ abundances \citep{kulyk2021}. CO was not detected, but CO$_2$ was detected with the AKARI telescope, so the CO/CO$_2$ ratios obtained are upper limits, with the most conservative (highest) value being $<$0.45 \citep{oot2012}, consistent with a CO$_2$-dominated coma.

\subsection{C/2006 W3 Christensen}
\label{subsec:2006W3}

This distantly active comet never approached closer than $\sim$ 3 au of the Sun during its perihelion passage and was found to be rich in molecules more volatile than water and the distant CO production was proposed to be largely released during the crystallization of amorphous water ice near the nucleus surface \citep{bockelee2010,prialnik2004}. CO and CO$_2$ emission was detected by AKARI with values ranging from Q$_{CO}$/Q$_{CO_2}$ = 2.3 to 3.5. This high ratio was attributed to being at least partly due to insufficient sublimation of CO$_2$ at relatively large distances of 3.13 and 3.66 au from the Sun \citep{oot2012}.

\subsection{C/2007 N3 Lulin}
\label{subsec:2007N3}

C/2007 N3 is a dynamically new comet with two near polar jets and a morphology that is different from that of the water outgassing, and thus was proposed to have either CO$_2$ or CO as the main drivers of activity \citep{gibb2012,bair2018}. Given that Q$_{CO_2}$ $>>$ Q$_{CO}$ from the AKARI measurements, it seems that the jets were most likely driven by CO$_2$ outgassing. Comparison with water production rates also shows that CO was present in very small amounts with the highest abundance ratio reaching 2.2\%. In this coma very low relative to H$_2$O \citep{oot2012, Gibb2012b}, see Table \ref{tab:observations}, and Table \ref{tab:supobservations}.

\subsection{C/2009 P1 Garradd}
\label{subsec:2009P1}

The HRI-IR detected for both CO and CO$_2$ emission in C/2009 P1 during its extended EPOXI mission \citep{ahearn2011, feaga14}. 

The CO/CO$_2$ ratio was 7.44, corresponding to a CO-dominant coma. C/2009 P1 Garradd's active fraction of the nucleus exceeded 50 \% \citep{bois2013} and the CO production significantly increased after perihelion \citep{bodewits2014, feaga14}. 

\subsection{C/2012 S1 ISON}
\label{subsec:2012S1}

The CO$_2$ production rate for comet C/2012 S1 was calculated using Spitzer 4.5 $\mu$m combined CO+CO$_2$ images \citep{lisse2013} and an independently modeled CO production rate that agreed with estimates from CO from HST and a marginal CO detection at mm-wavelengths. Modeling showed that CO$_2$ was an unlikely match to the Spitzer excess emission \citep{Meech13} and instead the activity was probably driven by CO. We used this CO value to extract a CO$_2$ production rate from the Spitzer measurements.


\subsection{C/2016 R2 PanSTARRS}
\label{subsec:2016R2}


C/2016 R2 pre-perihelion observations were taken from Arizona Radio Observatory 10-m SMT (CO J=2-1) on February 13, 2018 at 5:16 UT and with Spitzer IRAC on February 12, 2018 at 18:22 UT (CO$_2$+CO), with a difference of $\sim$ 11 hours between them. The post-perihelion observations were taken on June 12, 2019 with the ARO SMT and on June 10, 2019 with Spitzer IRAC so the date between them June 11, 2019, was used to represent the data. CO detections observed pre-perihelion taken a couple weeks before by the IRTF iShell were in agreement with the CO obtained with the SMT \citep{mckay2019}, but since we were trying to use the measurements that were taken the closest together in time we use the CO from the SMT.

Unlike other comets that are mainly composed of H$_2$O, the dominant volatiles in comet C/2016 R2 are N$_2$ and CO \citep{mckay2019}. Interestingly, despite extraordinarily high fractional abundances of CO and CO$_2$ with water, the CO/CO$_2$ (determined by Equations \ref{eq:spitzer} and \ref{eq:gfactor}) ratios are more similar to what other Oort Cloud comets had at similar distances (Figure \ref{fig:indivratios}). Due to its distinct molecular composition, it is believed that C/2016 R2 formed further out than where H$_2$O rich comets formed and the likelihood of observing a similar comet to C/2016 R2 is low \citep{mousis2021}. The factor of 3.6 increase in the CO/CO$_2$ ratio from pre-perihelion to post-perihelion looks like fellow OCC, C/1995 O1 Hale-Bopp. Not only is this comet CO rich in the coma, it is very H$_2$O poor, and some have wondered whether it could have formed beyond the N$_2$ ice line or be a fragment from a TNO \citep{wie18, biver2018}.

C/2016 R2 also has the highest fractional abundances of CO/H$_2$O and CO$_2$/H$_2$O distinct from composition from all other comets. Even the interstellar comet, 2I/Borisov, only had a CO/H$_2$O ratio of 1.425 at 2.12 au while C/2016 R2's CO/H$_2$O ratio was 177.42 at 2.76 au.

\section{Summary and Conclusions}
\label{sec:Conclusion}

We compiled a comprehensive survey of published contemporaneous CO and CO$_2$ production rates in 25 comets, and we include water production rates and nucleus diameters when available. We supplemented this analysis with additional measurements of water, CO, and CO$_2$. From this dataset we analyzed key production rate ratios, including the volatile C/O ratio, and searched for possible correlations with heliocentric distance, orbital families, and nucleus surface area. Key findings are listed below, including some interesting trends, which can be tested with future observations.

\begin{enumerate}

\item We infer Q$_{CO_2}$ = (5$\pm$3)x10$^{26}$ mol s$^{-1}$ for 29P using contemporaneous data from the Spitzer and the IRAM 30-m telescopes. This value indicates that CO$_2$ is present in the Centaur's coma at the level of $\sim$ 1\% of CO and 5\% of H$_2$O. 29P clearly has the highest CO/CO$_2$ mixing ratio in this study. It is not clear how much of this difference between CO and CO$_2$ release is due to chemical composition of the nucleus and how much can be attributed to preferential CO release over the less volatile CO$_2$ at $\sim$ 6 au from the Sun. 

\item Collectively, approximately half of the 25 comets had more CO$_2$ than CO in their coma, one third were CO-dominated, and about a tenth had comparable amounts of CO and CO$_2$. 

\item Eight of the nine JFCs had CO$_2$-dominated comae, which may be partly explained by preferential loss of CO vs. CO$_2$ over millenia. 

\item All six comets on their first trip from the Oort Cloud to the inner solar system (dynamically new) produced more CO$_2$ than CO when within 2.5 au of the Sun, while the remaining eight comets, which were on their second or later trip, produced more CO at this distance. DN comets produce more CO$_2$ relative to CO than the other Oort Cloud Comets that have made multiple passes through the inner solar system. There may also be a trend where the Q$_{CO}$/Q$_{CO_2}$ ratio increases with dynamical age, as first pointed out by \citet{Ahearn2012}. These results are inconsistent with models that predict that the least processed comets should have more CO than CO$_2$ 
and raises questions about comet formation and evolution models. 
A low CO/CO$_2$ production rate ratio for dynamically new comets may instead be explained by a model that includes galactic cosmic rays processing that alters the ice composition down to meters below the surface, resulting in CO-depletion with respect to CO$_2$ \citep{gronoff2020, maggiolo2020}. This highly processed layer is then eroded during the first perihelion passage, revealing relatively fresher surface layers with pockets of CO ice that are liberated during subsequent passages. 

\item All comae observed beyond 3.5 au were dominated by CO. This could be due to the higher volatility of CO, but a selection effect due to the small sample size of active comets beyond 3.5 au could also be a contributing factor. Based on the results of this study, when independent measurements are not possible for both species for an active comet beyond 3.5 au, it may be appropriate to assume that the dominant volatile is CO rather than CO$_2$.

\item Given the possible trend of increasing Q$_{CO}$/Q$_{CO_2}$ with larger heliocentric distance, and the lack of JFC measurements of these species beyond $\sim$ 3.5 au, there may be a selection effect which leads to the lower Q$_{CO}$/Q$_{CO_2}$ in JFCs. Many additional measurements of Q$_{CO}$/Q$_{CO_2}$ in JFCs at larger distances, especially beyond 4 au, are needed to determine the contributions due to thermal heating and chemical composition of the nucleus as primary contributors for the apparent difference between the JFC and OCC comae.

\item If the trend of Q$_{CO_2}$ $\ge$ Q$_{CO}$ for dynamically new comets holds (Section \ref{sec:dynamicalage}), then we anticipate that the incoming DN comet C/2017 K2 (PanSTARRS) should produce more CO$_2$ than CO when it is near perihelion.

\item For comets within 2.5 au, the CO/H$_2$O production rate ratio has a median of 3 \textpm{ 1\%}, with a range from 0.3\% to 26\%. For CO$_2$/H$_2$O, the median is higher, 12 \textpm{ 2\%} with a range from 2\% to 30\%. The production rate ratio for CO may change slope at $\sim$ 2.5-3.0 au, where water ice sublimation noticeably changes its efficiency.

\item For all comets from 0.7 to 4.6 au, Q$_{CO_2}$/Q$_{H_2O}$ shows a much tighter correlation (less scatter) with respect to heliocentric distance than does Q$_{CO}$/Q$_{H_2O}$. This may mean that the production of CO$_2$ and CO has significantly different mechanisms over this range, and that CO$_2$ outgassing may be more intimately tied to water production than CO is. In contrast, the fractional CO production rate has a much stronger response to solar heating than does CO$_2$, which is consistent with CO being much more volatile.

\item The median production rate ratios of (Q$_{CO}$ + Q$_{CO_2}$)/Q$_{H_2O}$ for all comets within 2.5 au is 18 \textpm{ 4\%}, although there is some substantial deviation from this value for a few comets. This indicates that the total amounts of CO and CO$_2$ produced within 2.5 au is conserved for most comets and represents $\sim$ 20\% of the total volatile component, as proposed by \citet{Ahearn2012, 2021Lisse}. This 
may support the theories that comets retain a strong amount of their natal composition.

\item When observed over the heliocentric distance range of 1 au to 2 au, most Oort Cloud comets produce significantly more CO per surface area than JFCs. No distinction is seen for CO$_2$ between the two groups of comets. 

\item We computed the carbon-to-oxygen ratio for the gaseous component using contemporaneous production rates of CO, CO$_2$, and H$_2$O in 18 comets, the largest sources of carbon and oxygen in comae. We find C/O$_{average}$ $\sim$ 15\% and C/O$_{median}$ $\sim$ 13\%, which is consistent with most comets forming within the CO snow line (i.e., C/O $<$ 0.2) \citep{Seligman2022}. A notable exception is the anomalous C/2016 R2, with C/O $\sim$ 0.9, which can be explained by forming farther out, possibly beyond the CO snow line. We see no distinction between the JFCs and OCCs with respect to C/O ratios. This may be explained by models which predict that both Kuiper Belt and Oort Cloud reservoirs have similar origins and that they both formed within the region of the giant planets (i.e. 5-30 au).

\end{enumerate}

\subsection{Large Tables}
\begin{longrotatetable}
\begin{deluxetable*}{lccrrrrrrrrc}
\renewcommand{\arraystretch}{0.85}
\label{tab:observations}
\tabletypesize{\scriptsize}
\tablecaption{Contemporaneous CO, CO$_2$, and H$_2$O Measurements in Comets} 
\tablecolumns{12}
\tablewidth{0pt}
\tablehead{
\colhead{Comet} &
\colhead{Date} &
\colhead{R$_{Helio}$} &
\colhead{True Anomaly} &
\colhead{Q$_{CO}$} &
\colhead{Q$_{CO_2}$} &
\colhead{Q$_{H_2O}$} &
\colhead{ Q$_{CO}$/Q$_{CO_2}$} & 
\colhead{ Q$_{CO}$/Q$_{H_2O}$} &  
\colhead{ Q$_{CO_2}$/Q$_{H_2O}$} &  
\colhead{1/$a_0$}&
\colhead{Ref.}\\      
\colhead{} & 
\colhead{(YYYY/MM/DD)} &
\colhead{(AU)} &
\colhead{(degrees)} &
\colhead{$10^{26}$ sec$^{-1}$} &
\colhead{$10^{26}$ sec$^{-1}$} &
\colhead{$10^{26}$ sec$^{-1}$} &
\colhead{} & 
\colhead{} & 
\colhead{} & 
\colhead{$10^{-4}$ AU$^{-1}$} & 
\colhead{}
}
\startdata
     1P/Halley & 1986/03/06 & 0.79 & 61 & 500.00 & 180.00 & 4500.0 & 2.78 & 0.110 & 0.04 & - &2\\
     9P/Tempel 1 & 2005/07/03 & 1.51 & 359 & 5.00 & - & - & - & - & - & - & 6 \\
     9P/Tempel 1 & 2005/07/04 & 1.51 & 359 & - & 3.20 & 46.0 \footnote{Water from July 4 pre-impact.} & 1.88 & 0.130 & 0.07 & - &5 \\
     21P/G-Z & 2018/10/03 & 1.07 & 28 & - & - & - & - & - & 0.11 & - &21\\
     21P/G-Z & 2018/10/10 & 1.10 & 36 & 2.55 & - & 203.7 & 0.11 & 0.013 & - & - &20\\
     22P/Kopff & 2009/04/22 & 1.61 & 340 & $<$1.65 & 10.96 & 59.4 & $<$0.15 & $<$0.030 & 0.18 & - & 1\\
     29P/S-W 1 & 2009/11/18 & 6.18 & 135 & 291.5 & $<$3.5 & 67.8 & $>$83.2 & 4.630 & $<$0.05 & - &1\\
     46P/Wirtanen & 2019/01/08 & 1.11 & 29 & - & 8.55 & - & $<$0.04 & - & 0.15 & - &11 \\
     46P/Wirtanen & 2019/01/11 & 1.13 & 33 & $<$0.31 & - & 57.0 & $<$0.04 & $<$0.005 & - & - &11 \\
     67P/C-G & 2014/08/23 & 3.50 & 229 & 0.21 & 0.07 & 0.5 & 3.06 & 0.450 & 0.15 & - &12 \\
     67P/C-G & 2015/08/13 & 1.24 & 359 & 4.78 & 6.26 & 158.1 & 0.76 & 0.030 & 0.04 & - &12 \\ 
     67P/C-G & 2016/01/02 & 2.03 & 89 & 0.36 & 2.71 & 11.5 & 0.13 & 0.030 & 0.24 & - & 12 \\
     81P/Wild 2& 2009/12/14 & 1.74 & 320 & $<$2.24 & 8.63 & 60.0 & $<$0.26 & $<$0.040 & 0.14 & - & 1\\
     88P/Howell & 2009/07/03 & 1.73 & 294 &$<$2.72 & 8.65 & 33.5 & $<$0.31 & $<$0.080 & 0.26 & - &1\\
     103P/Hartley 2& 2010/11/02 & 1.06 & 8 & - & 20.00 & 100.0 & 0.01 & 0.003 & 0.20 & - &4 \\
     103P/Hartley 2& 2010/11/04& 1.06 & 8 & 0.26 & - & - & - & - & - & - & 3 \\
     144P/Kushida & 2009/04/18 & 1.70 & 53 &$<$1.14 & 5.57 & 37.6 & $<$0.20 & $<$0.030 & 0.15 & - &1\\
     C/1979 Y1 & 1980/01/10 & 0.71 & 56 & 51.00 & 51.00 & 1460.0 & 1.00 & 0.030 & 0.03 & 232.7 &13,15\\
     C/1989 X1 & 1990/05/09 & 0.83 & 99 & 17.00 & 21.00 & 1020.0 & 0.81 & 0.020 & 0.02 & 0.3 &13,15\\
     C/1990 K1 & 1990/08/26 & 1.38 & 291 & 101.00 & 170.00 & 2480.0.0 & 0.59 & 0.040 & 0.07 & 0.7 &13,15\\
     C/1990 K1 & 1990/09/18 & 1.13 & 311 & 168.00 & 268.00 & 2010.0 & 0.63 & 0.080 & 0.13 & 0.7 &13,15\\
     C/1995 O1 & 1996/04/27 & 4.58 & 233 &700.00 & 130.00 & 131.0 & 5.38 & 5.340 & 0.99 & 38.4 &7,16\\
     C/1995 O1 & 1996/09/26 & 2.93 & 248 &2300.00 & 740.00 & 3300.0 & 3.11 & 0.700 & 0.22 & 38.4 &7\\
     C/1995 O1 & 1997/12/29 & 3.89 & 122 &1500.00 & 350.00 & 280.0 & 4.29 & 5.360 & 1.25 & 38.4 &7\\
     C/1995 O1 & 1998/04/06 & 4.90 & 129 &3000.00 & 300.00 & - & 10.00 & - & - & 38.4 &7\\
     C/1996 B2 & 1996/06/02 & 0.94 & 170 & 180.00 & 60.00 & 1530 & 2.83 & 0.111 & 0.04 & 15.5 &17\\
     C/2006 OF2& 2008/09/16 & 2.43 & 0 & $<$2.52 & 13.97 & 61.2 & $<$0.18 & $<$0.040 & 0.23 & 0.2 & 1\\
     C/2006 OF2& 2009/03/28 & 3.20 & 59 &$<$4.46 & 9.90 & 17.0 & $<$0.45 & $<$0.260 & 0.58 & 0.2 & 1\\
     C/2006 Q1& 2008/06/03 & 2.78 & 351 &$<$3.67 & 16.19 & 36.3 & $<$0.23 & $<$0.100 & 0.45 & 0.4 &1\\
     C/2006 W3 & 2008/12/21 & 3.66 & 315 &299.30 & 84.59 & 83.0 & 3.54 & 3.610 & 1.02 & 3.7 & 1\\
     C/2006 W3 & 2009/06/16 & 3.13 & 355 &197.70 & 84.66 & 201.0 & 2.34 & 0.980 & 0.42 & 3.7 & 1\\
     C/2007 N3 & 2009/02/05 & 1.28 & 26 & $<$8.67 & 48.48 & 409.1 & $<$0.18 & $<$0.020 & 0.12 & 0.3 & 1\\
     C/2007 N3 & 2009/03/30 & 1.70 & 64 &$<$2.86 & 22.42 & 223.8 & $<$0.13 & $<$0.010 & 0.10 & 0.3 & 1\\
     C/2007 Q3 & 2009/03/03 & 3.29 & 292 &$<$3.97 & 6.95 & 39.8 & $<$0.57 & $<$0.100 & 0.17 & 0.3 & 1\\
     C/2008 Q3 & 2009/07/05 & 1.81 & 7 &29.30 & 31.04 & 112.0 & 0.94 & 0.260 & 0.28 & 2.5 & 1\\
     C/2009 P1 & 2012/03/26 & 2.00 & 56 & 290.00 & 39.00 & 460.0 & 7.44 & 0.630 & 0.08 & 4.2 & 9\\
     C/2012 S1 & 2013/06/13 \footnote{This is with the date for Spitzer and the modeled CO.}& 3.34 & 187 & 7.00 & 1.21 & $<$100.0 \footnote{Water used from May 9$^{th}$}& 5.79 & * & * & 1.2 & 10,19\\
     C/2016 R2 & 2018/02/12 & 2.76 & 332 & - & 90.50 & - & - & - & - & 11.7 & 8\\
     C/2016 R2 & 2018/02/13 & 2.76 & 333 & 550.00 & - & - & 6.08 & - & - & 11.7 & 8\\
     C/2016 R2 & 2018/02/21 &2.73 & 335 & - & - & 3.1 \footnote{Water from DCT LMI via (OH) this is the closest water detection to the dates CO and CO$_2$ were observed. Water is from Feb.21, 2018.} & - & 177.420 & 29.19 & 11.7 & 8\\
     C/2016 R2 & 2019/06/10 & 4.72 & 84 & - & 6.21 & - & 19.00 & - & - & 11.7 & 14\\
     C/2016 R2 & 2019/06/12 &4.73 & 84 & 118.00 & - & - & - & - & - & 11.7 & 18\\
\enddata
\tablerefs{ 1: \citet{oot2012}, 2: \citet{combes1988}, 3: \citet{weaver2011}, 4: \citet{ahearn2011}, 5: \citet{feaga2007}, 6: \citet{feldman2006}, 7: \citet{crovisier1999}, 8: \citet{mckay2019}, 9: \citet{feaga14},
10: \citet{lisse2013},
11: \citet{Mckay21},
12: \citet{Combi2020},
13: \citet{feldman1997},
14: private communication with Adam McKay (Spitzer CO proxy),
15: \citet{tozzi1998}, 16: \citet{colom1997} OH production given as a proxy for H$_2$O, 17: \citet{mcphate1999} H$_2$O is derived from the OH band, 18: \citet{wier2019}, 19: \citet{Meech13}, 20:\citet{Roth2020}, 21: \citet{Shinnaka2020} The 1/$a_0$ values are from the MPC database and are used to construct Figure \ref{fig:dynageOCC}. *: The water production is an upper limit the production rate ratios.
}
\end{deluxetable*}

\end{longrotatetable}

\begin{longrotatetable}
\begin{deluxetable*}{lccrrrrc}
\renewcommand{\arraystretch}{0.85}
\tablecaption{Additional CO, CO$_2$, and H$_2$O Measurements in Comets}
\tabletypesize{\scriptsize}
\label{tab:supobservations}
\tablecolumns{8}
\tablewidth{0pt}
\tablehead{
\colhead{Comet} &
\colhead{Date} &
\colhead{R$_{Helio}$} &
\colhead{Q$_{H_2O}$} &
\colhead{Q$_{CO}$} &
\colhead{Q$_{CO_2}$} &
\colhead{ Q$_{CO}$/Q$_{H_2O}$} &  
\colhead{Ref.}\\      
\colhead{} & 
\colhead{(YYYY/MM/DD)} &
\colhead{(AU)} &
\colhead{$10^{26}$ sec$^{-1}$} &
\colhead{$10^{26}$ sec$^{-1}$} &
\colhead{$10^{26}$ sec$^{-1}$} &
\colhead{or Q$_{CO_2}$/Q$_{H_2O}$} & 
\colhead{}
}
\startdata
     19P/Borrelly & 2008/12/30 & 2.19 & 6.4 & - & 1.6 & 0.250 & 16\\ 
     64P/Swift–Gehrels & 2009/11/23 & 2.27 & 6.9 & - & 1.5 & 0.220 & 16\\
     103P/Hartley 2& 1997/12/31 & 1.04 & - & - & 12.0 & - & 1 \\
     103P/Hartley 2& 1998/01/01 & 1.04 & 124.0 & - & - & 0.097 & 1 \\
     116P/Wild 4 & 2009/05/16 & 2.22 & 15.8 & - & 0.9 & 0.056 & 16\\
     118P/Shoemaker–Levy 4 & 2009/09/08 & 2.18 & 10.0 & - & 3.0 & 0.300 & 16\\
     157P/Tritton & 2009/12/30 & 1.48 & 4.5 & - & 0.3 & 0.069 & 16\\
     C/2007 G1 (LINEAR) & 2008/08/20 & 2.80 & 18.4 & - & 4.2 & 0.230 & 16\\
     8P/Tuttle & 2008/01/27 & 1.03 & 544.0 & 2.4 & - & 0.004 & 24\\ 
     8P/Tuttle & 2007/12/23 & 1.15 & 212.8 & $<$0.8 & - & $<$0.004 & 25\\ 
     21P/Giacobini–Zinner & 1998/10/02 & 1.25 & 320.0 & 33.0 & - & 0.100 & 27\\
     21P/Giacobini–Zinner & 2018/07/30 & 1.17 & 231.6 & 5.5 & - & 0.024 & 28\\ 
     21P/Giacobini–Zinner & 2018/10/07 & 1.09 & 258.3 & 3.7 & - & 0.014 & 28\\ 
     41P/Tuttle–Giacobini–Kresák & 2017/04/26 & 1.06 & 50.0 & $<$8.5 & - & $<$0.170 & 28\\
     45P/Honda–Mrkos–Pajdušáková & 2017/01/08 & 0.56 & 372.0 & 1.9 & - & 0.005 & 31\\
     73P/Schwassmann–Wachmann 3 & 2006/05/27 & 0.95 & 131.0 & 0.6 & - & 0.005 & 26\\
     153P/Ikeya–Zhang & 2002/04/20 & 0.89 & 2150.0 & 154.0 & - & 0.072 & 23\\ 
     C/1995 O1 (Hale-Bopp) & 1997/01/27 & 1.49 & * & * & - & 0.267 & 22 \\
     C/1995 O1 (Hale-Bopp) & 1997/03/01 & 1.06 & * & * & - & 0.271 & 22 \\
     C/1995 O1 (Hale-Bopp) & 1997/04/09 & 0.93 & * & * & - & 0.276 & 22 \\
     C/1995 O1 (Hale-Bopp) & 1997/05/01 & 1.06 & * & * & - & 0.280 & 22 \\
     C/1996 B2 (Hyakutake) & 1996/03/24 & 1.06 & 2540.0 & - & - & - & 20\\
     C/1996 B2 (Hyakutake) & 1996/03/24 & 1.06 & - & 440.0 & - & 0.170 & 21\\ 
     C/1996 B2 (Hyakutake) & 1996/04/12 & 0.64 & 3990.0 & - & - & - & 20\\
     C/1996 B2 (Hyakutake) & 1996/04/12 & 0.64 & - & 803.0 & - & 0.200 & 21\\
     C/1999 H1 (Lee) & 1999/08/20 & 1.06 & 1260.0 & 23.0 & - & 0.018 & 19\\
     C/1999 S4 (LINEAR) & 2000/07/13 & 0.81 & 446.0 & 2.0 & - & 0.004 & 18\\
     C/1999 T1 (McNaught-Hartley) & 2001/01/13 & 1.3 & 820.0 & 140.0 & - & 0.170 & 17\\
     C/2000 WM1 (LINEAR) & 2001/11/25 & 1.32 & 177.1 & 0.9 & - & 0.005 & 15\\
     C/2001 A2 (LINEAR) & 2001/07/10 & 1.17 & - & 16.6 & - & - & 13\\
     C/2001 A2 (LINEAR) & 2001/07/10 & 1.17 & 430.0 & - & - & 0.039 & 14\\
     C/2001 Q4 (NEAT) & 2004/04/26 & 1.02 & 2000.0 & 176.0 & - & 0.088 & 23\\
     C/2004 Q2 (Machholz) & 2004/11/29 & 1.48 & 1253.0 & 63.5 & - & 0.051 & 12\\
     C/2006 M4 (SWAN) & 2006/11/07 & 1.08 & 1756.0 & 8.7 & - & 0.005 & 11\\
     C/2006 P1 (McNaught) & 2007/01/27 & 0.55 & 17400.0 & 341.0 & - & 0.020 & 10\\
     C/2007 N3 (Lulin) & 2009/02/01 & 1.26 & 2013.0 & 43.6 & - & 0.022 & 9\\
     C/2007 W1 (Boattini) & 2008/07/10 & 0.90 & 122.3 & 5.5 & - & 0.045 & 8 \\
     C/2009 P1 (Garradd) & 2011/09/21 & 2.01 & 1410.0 & 67.0 & - & 0.048 & 7 \\
     C/2009 P1 (Garradd) & 2011/10/10 & 1.85 & 1640.0 & 103.0 & - & 0.063 & 7\\
     C/2009 P1 (Garradd) & 2012/01/25 & 1.62 & 1050.0 & 164.0 & - & 0.160 & 7\\
     C/2009 P1 (Garradd) & 2012/02/27 & 1.69 & 1000.0 & 196.0 & - & 0.196 & 7\\
     C/2010 G2 (Hill) & 2012/01/10 & 2.51 & 61.0 & 47.0 & - & 0.770 & 6 \\
     C/2012 S1 (ISON) & 2013/11/17 & 0.53 & 1827.0 & 22.4 & - & 0.012 & 4 \\
     C/2012 F6 (Lemmon) & 2013/03/31 & 0.75 & 4590.0 & 195.0 & - & 0.042 & 5 \\
     C/2013 R1 (Lovejoy) & 2013/10/24 & 1.34 & 153.2 & 15.2 & - & 0.099 & 2 \\
     C/2013 R1 (Lovejoy) & 2013/12/10 & 0.84 & * & * & - & 0.123 & 3 \\
     C/2020 F3 (NEOWISE) & 2020/07/20 & 0.56 & 4413.0 & 82.8 & - & 0.019 & 29\\
     C/2020 F3 (NEOWISE) & 2020/08/01 & 0.83 & 1176.5 & 30.0 & - & 0.026 & 29\\
     2I/Borisov$_{average}$ & 2019/(12/19-22), 01/13/2020 & 2.12 (median) & ** & ** & - & 1.425 (median) & 30 \\
\enddata 
\tablerefs{ 1: \citet{crovisier1999b}, 2: \citet{paganini2014}, 3: \citet{dello2016}, 4: \citet{disanti2016}, 5: \citet{paganini2014b}, 6: \citet{Kawakita2014}, 7: \citet{mckay2015}, 8:\citet{Villanueva2011}, 9: \citet{Gibb2012b}, 10: \citet{dello2009}, 11: \citet{DiSanti2009}, 12: \citet{bonev2009}, 13: \citet{mageesauer2008}, 14: \citet{dello2005}, 15: \citet{radeva2010}. 16: \citet{oot2012}, 17: \citet{mumma2001}, 18: \citet{mumma2001c}, 19: \citet{mumma2001b}, 20: \citet{dello2002}, 21: \citet{disanti2003}, 22: \citet{disanti2001}, 23: \citet{Lupu2007}, 24: \citet{bohnhardt2008}, 25: \citet{bonev2008}, 26: \citet{disanti2007}, 27: \citet{mumma2000}, 28: \citet{Faggi2019}, 29: \citet{faggi2021}, 30: \citet{bodewits2020}, 31: \citet{disanti2017}, * = Comets for which only the CO/H$_2$O ratio was given (rather than production rates, ** = This ratio was obtained as the median of the rage of ratios from the dates given.)
}
\end{deluxetable*}
\end{longrotatetable}

\begin{longrotatetable}
\begin{deluxetable*}{lccrrrc}
\renewcommand{\arraystretch}{0.85}
\tablecaption{CO and Diameter Measurements} 
\label{tab:surfobsd}
\tabletypesize{\scriptsize}
\tablecolumns{7}
\tablewidth{0pt}
\tablehead{
\colhead{Comet} &
\colhead{Date} &
\colhead{R$_{Helio}$ for Q$_{CO}$} &
\colhead{Q$_{CO}$} &
\colhead{Diameter} &
\colhead{ Q$_{CO}$/D$^2$} &  
\colhead{References}\\      
\colhead{} & 
\colhead{(YYYY/MM/DD)} &
\colhead{(AU)} &
\colhead{$10^{26}$ sec$^{-1}$} &
\colhead{km} &
\colhead{$10^{26}$ sec$^{-1}$ km$^{-2}$ } & 
\colhead{}
}
\startdata
     1P/Halley & 1986/03/13 & 0.9 & 281.00 & 11.0 & 2.320 & 1,2 \\
     2P/Encke & 2003/11/05-06 & 1.2 & 0.57 & 4.8 & 0.025 & 20,2\\
     8P/Tuttle & 2008/01/27 & 1.0 & 2.43 & 6.2 & 0.063 & 12,13\\
     9P/Temple 1 & 2005/07/03 & 1.5 & 5.00 & 5.95 & 0.206 & 14,45\\
     21P/Giacobini–Zinner & 2018/10/07 & 1.1 & 3.72 & 2.0 & 1.500 & 18,2\\
     29P/Schwassmann–Wachmann 1 & 2016/02/28 & 6.0 & 593.00 & 60.0 & 0.165 & 30,44 \\
     41P/Tuttle–Giacobini–Kresák & 2017/04/01-02 & 1.1 & 3.80 & 1.4 & 1.940 & 16,2 \\
     45P/Honda–Mrkos–Pajdušáková & 2017/01/08 & 0.6 & 1.87 & 1.6 & 0.730 & 17,2\\
     46P/Wirtanen & 2019/01/11 & 1.1 & $<$0.31 & 1.1 & 0.250 & 39,40\\
     67P/Churyumov–Gerasimenko & 2014/08/23 & 3.5 & 0.21 & 3.44 & 0.017 & 26,31\\
     67P/Churyumov–Gerasimenko & 2015/01/18 & 2.5 & 0.04 & 3.44 & 0.004 & 26,31\\
     67P/Churyumov–Gerasimenko & 2015/10/31 & 1.6 & 2.27 & 3.44 & 0.192 & 26,31\\
     77P/Longmore* & 2010/05/05 & 3.0 & 3.31 & 8.2 & 0.049 & 25,22\\
     77P/Longmore* & 2010/08/16 & 3.3 & 0.37 & 8.2 & 0.006 & 24,22\\
     81P/Wild 2* & 2010/04/12 & 1.7 & 0.37 & 8.43 & 0.005 & 24,45\\
     88P/Howell & 2009/07/03 & 1.7 & $<$2.72 & 4.4 & $<$0.140 & 21,22\\
     94P/Russell 4* & 2010/07/13 & 2.3 & 0.35 & 5.2 & 0.013 & 24,23\\
     103P/Hartley 2 & 2010/11/04 & 1.1 & 0.26 & 1.27 & 0.161 & 19,45\\
     174P/Echeclus & 2016/(05/28– 06/09) & 6.1 & 7.700 & 83.6 & 0.0011 & 4,41 \\
     (2060) Chiron & 1995/06/10-12 & 8.5 & 130.00 & 218.0 & 0.003 & 5,6 \\
     342842 & 2011/10/28 & 6.6 & $<$13.18 & 67.0 & $<$0.003 & 7,8 \\
     95626 & 2002/05/22 & 21.1 & $<$64.57 & 237.0 & $<$0.001 & 7,9 \\
     (2060) Chiron & 1999/06/25 & 9.5 & $<$60.26 & 218.0 & $<$0.001 & 7,6\\
     (10199) Chariklo & 1999/02/28 & 13.5 & $<$83.18 & 248.0 & $<$0.001 & 7,6\\
     C/1996 B2 (Hyakutake) & 1996/02/10 & 1.9 & 124.00 & 4.2 & 7.030 & 10,11\\
     C/1996 B2 (Hyakutake) & 1996/03/24 & 1.1 & 517.00 & 4.2 & 29.300 & 10,11\\
     C/1996 B2 (Hyakutake) & 1996/04/28 & 0.3 & 1080.00 & 4.2 & 61.200 & 10,11\\
     C/1999 S4 (LINEAR) & 2000/07/13 & 0.8 & 2.00 & 0.9 & 2.470 & 27,28\\
     C/2001 A2 (LINEAR) & 2001/07/10 & 1.2 & 16.60 & $<$3.0 & $>$1.840 & 32,33\\
     C/2006 OF2 (Broughton) & 2008/09/16 & 2.4 & $<$2.52 & 12.5 & $<$0.016 & 21,45\\
     C/2006 Q1 (McNaught)& 2008/06/03 & 2.8 & $<$3.67 & 16.2 & $<$0.014 & 21,45\\
     C/2006 W3 (Christensen) & 2008/12/21 & 3.7 & 299.30 & 43.75 & 0.156 & 21,45\\
     C/2006 W3 (Christensen) & 2009/06/16 & 3.1 & 197.70 & 43.75 & 0.103 & 21,45\\
     C/2007 G1 (LINEAR) & 2008/08/20 & 2.8 & $<$3.19 & 10.5 & $<$0.029 & 21,45\\
     C/2007 N3 (Lulin) & 2009/02/05 & 1.3 & $<$8.67 & 12.2 & $<$0.058 & 21,45\\
     C/2007 N3 (Lulin) & 2009/03/30 & 1.7 & $<$2.86 & 12.2 & $<$0.019 & 21,45\\
     C/2007 Q3 (Siding Spring) & 2009/03/03 & 3.3 & $<$3.97 & 18.3 & $<$0.012 & 21,45\\
     C/2008 Q3 (Garradd) & 2009/07/05 & 1.8 & 29.30 & 6.7 & 0.653 & 21,45\\
     C/2009 P1 (Garradd) & 2012/03/26 & 2.0 & 290.00 & 27.0 & 0.398 & 46,45\\
     C/2010 G2 (Hill) & 2012/01/10 & 2.5 & 47.00 & 8.0 & 0.731 & 47,45\\
     C/2012 S1 (ISON) & 2013/11/17 & 0.5 & 22.00 & $<$1.3 & $>$14.080 & 34,35\\
     C/2014 UN271 (Bernardinelli–Bernstein) & 2020/11/26-28 & 20.9 & $<$200.00 & 119.0 & $<$0.014 & 42,43\\
     C/2016 S1 (PanSTARRS)* & 2016/12/01 & 2.6 & 3.67 & 5.2 & 0.140 & 3 \\
     C/2017 K2 (PanSTARRS) & 2021/04/01 & 6.7 & 16.00 & $<$18.0 & $>$0.050 & 36 \\
     2I/Borisov & 2019/12/11 & 2.1 & 7.50 & 0.4 $<$ D $<$1.0 & 46.900 $>$ P $>$7.500 & 37, 38\\
\enddata
\tablerefs{ 1: \citet{feldman1997}, 2: \citet{lamy2004}, 3: \citet{Rosser2018}, 4: \citet{Wierzchos2017}, 5: \citet{womack1999}, 6: \citet{Fornasier2013}, 7: \citet{drahus2017}, 8: \citet{bauer2013b}, 9: \citet{duffard2014}, 10: \citet{biver1999b}, 11: JPL Horizons Small-Body Database, 12: \citet{bohnhardt2008}, 13: \citet{harmon2010}, 14: \citet{feldman2006}, 15: \citet{ahearn2005}, 16: \citet{wierzchos2017b}, 17: \citet{disanti2017}, 18: \citet{Faggi2019}, 19: \citet{weaver2011}, 20: \citet{radeva2013}, 21: \citet{oot2012}, 22: \citet{Scotti1994}, 23: \citet{snodgrass2008}, 24: \citet{reach13}, 25: \citet{bau15}, 26: \citet{Combi2020}, 27: \citet{mumma2001c}, 28: \citet{altenhoff2002}, 29: \citet{biver2008}, 30: \citet{Wierzchos2020} the highest (from outburst) and lowest (from quiescent) CO productions are plotted in Figure \ref{fig:subcurve1}, 31: \citet{sierks2015}, 32: \citet{mageesauer2008}, 33: \citet{nolan2006}, 34: \citet{disanti2016}, 35: \citet{delamere2013}, 36: \citet{yang2021}, 37: \citet{bodewits2020}, 38: \citet{Jewitt2020}, 39:\citet{Mckay21}, 40:\citet{boehnhardt2002}, 41:\citet{stansberry2008},
42:\citet{hui2022},
43:\citet{farnham2021}, 44: \citet{schambeau2015}, 45: \citet{Bauer2017}, 46: \citet{feaga14}, 47: \citet{Kawakita2014}
*Q$_{CO}$s were obtained by multiplying by the Q$_{CO_2}$ by 11.6, these values are treated as upper limits. 
Encke type comets are under the same color scheme as JFCs. P = Q$_{CO}$/D$^2$
}
\end{deluxetable*} 
\end{longrotatetable}

\begin{longrotatetable}
\begin{deluxetable*}{lccrrrc}
\renewcommand{\arraystretch}{0.85}
\tablecaption{CO$_2$ and Diameter Measurements} 
\label{tab:surfobsdCO2}
\tabletypesize{\scriptsize}
\tablecolumns{7}

\tablewidth{0pt}
\tablehead{
\colhead{Comet} &
\colhead{Date} &
\colhead{R$_{Helio}$ for Q$_{CO_2}$} &
\colhead{Q$_{CO_2}$} &
\colhead{Diameter} &
\colhead{ Q$_{CO_2}$/D$^2$} &  
\colhead{References}\\      
\colhead{} & 
\colhead{(YYYY/MM/DD)} &
\colhead{(AU)} &
\colhead{$10^{26}$ sec$^{-1}$} &
\colhead{km} &
\colhead{$10^{26}$ sec$^{-1}$ km$^{-2}$ } & 
\colhead{}
}
\startdata
     1P/Halley & 1986/03/06 & 0.79 & 180.00 & 11.00 & 1.4900 & 11,12\\
     9P/Tempel 1 & 2005/07/04 & 1.51 & 3.20 & 5.95 & 0.0904 & 9,3\\
     19P/Borrelly & 2008/12/30 & 2.19 & 1.62 & 5.36 & 0.0564 & 1,3\\
     21P/Giacobini–Zinner & 2018/10/10 & 1.10 & 22.41 & 2.00 & 5.6000 & 20,21,12\\
     22P/Kopff & 2009/04/22 & 1.61 & 10.96 & 3.18 & 1.0800 & 1,2\\
     29P/Schwassmann–Wachmann 1 & 2009/11/18 & 6.18 & $<$3.22 & 60.40 & $<$0.0009 & 1,4\\
     46P/Wirtanen & 2019/01/11 & 1.13 & 8.55 & 1.12 & 6.8200 & 16,17\\
     64P/Swift–Gehrels & 2009/11/23 & 2.27 & 1.52 & 4.66 & 0.0700 & 1,3\\
     67P/Churyumov–Gerasimenko & 2014/08/23 & 3.50 & 0.07 & 3.44 & 0.0057 & 22,23\\
     67P/Churyumov–Gerasimenko & 2015/08/13 & 1.24 & 6.26 & 3.44 & 0.5290 & 22,23\\
     67P/Churyumov–Gerasimenko & 2016/01/02 & 2.03 & 2.71 & 3.44 & 0.2290 & 22,23\\
     81P/Wild 2 & 2009/12/14 & 1.74 & 8.63 & 8.43 & 0.1210 & 1,3\\
     88P/Howell & 2009/07/03 & 1.73 & 8.65 & 4.40 & 0.4470 & 1,5\\
     103P/Hartley 2 & 2010/11/02 & 1.06 & 20.00 & 1.27 & 12.4000 & 7,3\\
     116P/Wild 4 & 2009/05/16 & 2.22 & 0.89 & 8.25 & 0.0131 & 1,3\\
     118P/Shoemaker–Levy 4 & 2009/09/08 & 2.18 & 2.99 & 6.58 & 0.0691 & 1,2\\
     144P/Kushida* & 2009/04/18 & 1.70 & 5.57 & 1.62 & 2.1200 & 1,2 \\
     C/1995 O1 (Hale-Bopp) & 1996/04/27 & 4.58 & 130.00 & 60.00 & 0.0361 & 8,6\\
     C/1995 O1 (Hale-Bopp) & 1996/09/26 & 2.93 & 740.00 & 60.00 & 0.2060 & 8,6\\
     C/1995 O1 (Hale-Bopp) & 1997/12/29 & 3.89 & 350.00 & 60.00 & 0.0972 & 8,6\\
     C/1995 O1 (Hale-Bopp) & 1998/04/06 & 4.90 & 300.00 & 60.00 & 0.0833 & 8,6\\
     C/1996 B2 (Hyakutake) & 1996/06/02 & 0.94 & 63.00 & 4.80 & 2.7300 & 18,19\\
     C/2006 OF2 (Broughton) & 2008/09/16 & 2.43 & 13.97 & 12.54 & 0.0888 & 1,3\\
     C/2006 OF2 (Broughton) & 2009/03/28 & 3.20 & 9.90 & 12.54 & 0.0630 & 1,3\\
     C/2006 Q1 (McNaught)& 2008/06/03 & 2.78 & 16.19 & 16.23 & 0.0615 & 1,3\\
     C/2006 W3 (Christensen) & 2008/12/21 & 3.66 & 84.59 & 43.75 & 0.0442 & 1,3\\
     C/2006 W3 (Christensen) & 2009/06/16 & 3.13 & 84.66 & 43.75 & 0.0442 & 1,3\\
     C/2007 G1 (LINEAR) & 2008/08/20 & 2.80 & 4.17 & 10.50 & 0.0378 & 1,3\\
     C/2007 N3 (Lulin) & 2009/02/05 & 1.28 & 48.48 & 12.20 & 0.3260 & 1,3\\
     C/2007 N3 (Lulin) & 2009/03/30 & 1.70 & 22.42 & 12.20 & 0.1510 & 1,3\\
     C/2007 Q3 (Siding Spring) & 2009/03/03 & 3.29 & 6.95 & 18.30 & 0.0208 & 1,3\\
     C/2008 Q3 (Garradd) & 2009/07/05 & 1.81 & 31.04 & 6.70 & 0.6910 & 1,3\\
     C/2009 P1 (Garradd) & 2012/03/26 & 2.00 & 39.00 & 27.00 & 0.0535 & 10,3\\
     C/2012 S1 (ISON) & 2013/06/13 & 3.34 & 1.21 & $<$1.25 & $>$0.7740 & 13,14,15\\
\enddata
\tablerefs{ 1: \citet{oot2012}, 2: \citet{fernandez2013}, 3:\citet{Bauer2017}, 4: \citet{schambeau2015}, 5: \citet{Scotti1994}, 6: \citet{fernandez2002}, 7: \citet{ahearn2011}, 8: \citet{crovisier1999}, 9: \citet{feaga2007}, 10: \citet{feaga14}, 11: \citet{combes1988}, 12: \citet{lamy2004}, 13: \citet{lisse2013}, 14: \citet{Meech13}, 15: \citet{delamere2013}, 16: \citet{Mckay21}, 17: \citet{boehnhardt2002}, 18: \citet{mcphate1999}, 19: \citet{Lisse1999}, 20: \citet{Roth2020}, 21: \citet{Shinnaka2020}, 22: \citet{Combi2020}, 23: \citet{sierks2015}
*The smaller radius was used for comets which had multiple radii given.
}
\end{deluxetable*} 
\end{longrotatetable}


\begin{acknowledgments}
OHP acknowledges the LSSTC Data Science Fellowship Program, which is funded by LSSTC from NSF Cybertraining Grant \#1829740, the Brinson Foundation, and the Moore Foundation; her participation in the program has benefited from this work. This material is based on work supported by the National Science Foundation under grants AST-1615917 and AST-1945950. This material is based in part on work done by MW while serving at the National Science Foundation.

\end{acknowledgments}

\clearpage

\medskip




\bibliography{sample631}{}
\bibliographystyle{aasjournal}



\end{document}